\documentclass[onecolumn]{aastex631}

\usepackage{etoolbox}
\usepackage{amsmath}
\usepackage{amssymb}
\usepackage{tcolorbox}
\usepackage{savesym}
\usepackage[normalem]{ulem}
\usepackage{tabularx}
\usepackage{soul}
\usepackage{color}
\usepackage{graphicx}
\usepackage{bm}
\savesymbol{tablenum}
\usepackage{verbatim}
\usepackage{siunitx}
\restoresymbol{SIX}{tablenum}
\usepackage{academicons}
\usepackage[T1]{fontenc}
\definecolor{jblue}{RGB}{26, 102, 255}
\newcommand\ltsima{$\; \buildrel <\over\sim \;$}
\newcommand\simlt{\lower.5ex\hbox{\ltsima}}
\newcommand\gtsima{$\; \buildrel >\over\sim \;$}
\newcommand\simgt{\lower.5ex\hbox{\gtsima}}
\newcommand{\mathbold}[1]{\mbox{\boldmath $\bf#1$}}
\newcommand\piEbold{{\mathbold \pi_E}}

\renewcommand\hl[1]{#1}
\shorttitle{AAS\TeX\ A Jovian analogue orbiting a white dwarf star}
\shortauthors{Blackman et al.}

\begin{document}

\title{\Large A Jovian analogue orbiting a white dwarf star} 

%%AUTHORS

\author[0000-0001-5860-1157]{J.W. Blackman}
\affil{School of Natural Sciences, University of Tasmania,
Private Bag 37 Hobart, Tasmania 7001 Australia}
\affil{Sorbonne Universit\'es, UPMC Universit\'e Paris 6 et CNRS, 
UMR 7095, Institut d'Astrophysique de Paris, 98 bis bd Arago,
75014 Paris, France}

\author[0000-0003-0014-3354]{J.-P. Beaulieu}
\affil{School of Natural Sciences, University of Tasmania,
Private Bag 37 Hobart, Tasmania 7001 Australia}
\affil{Sorbonne Universit\'es, UPMC Universit\'e Paris 6 et CNRS, 
UMR 7095, Institut d'Astrophysique de Paris, 98 bis bd Arago,
75014 Paris, France}

\author[0000-0001-8043-8413]{D.P. Bennett}
\affil{Laboratory for Exoplanets and Stellar Astrophysics, NASA/Goddard Space Flight Center, Greenbelt, MD 20771, USA}
\affil{Department of Astronomy, University of Maryland, College Park, MD 20742, USA}

\author[0000-0002-3729-2663]{C. Danielski}
\affil{Instituto de Astrof\'isica de Andaluc\'ia (IAA-CSIC), Glorieta de la Astronom\'ia s/n, 18008 Granada, Spain}
\affil{UCL Centre for Space Exochemistry Data, Atlas Building, Fermi Avenue, Harwell Campus,  Didcot,  OX11 0QR, UK}
\affil{Sorbonne Universit\'es, UPMC Universit\'e Paris 6 et CNRS, UMR 7095, Institut d'Astrophysique de Paris, 98 bis bd Arago, 75014 Paris, France}

\author{C. Alard}
\affil{Sorbonne Universit\'es, UPMC Universit\'e Paris 6 et CNRS, 
UMR 7095, Institut d'Astrophysique de Paris, 98 bis bd Arago,
75014 Paris, France}

\author[0000-0003-0303-3855]{A.A. Cole}
\affil{School of Natural Sciences, University of Tasmania,
Private Bag 37 Hobart, Tasmania 7001 Australia}

\author[0000-0002-9881-4760]{A. Vandorou}
\affil{School of Natural Sciences, University of Tasmania,
Private Bag 37 Hobart, Tasmania 7001 Australia}

\author[0000-0003-2388-4534]{C. Ranc}
\affil{Laboratory for Exoplanets and Stellar Astrophysics, NASA/Goddard Space Flight Center, Greenbelt, MD 20771, USA}

\author[0000-0002-5029-3257]{S.K. Terry}
\affil{Department of Astronomy, University of California Berkeley, Berkeley, CA 94701, USA.}

\author{A. Bhattacharya}
\affil{Laboratory for Exoplanets and Stellar Astrophysics, NASA/Goddard Space Flight Center, Greenbelt, MD 20771, USA}
\affil{Department of Astronomy, University of Maryland, College Park, MD 20742, USA}

\author[0000-0002-8131-8891]{I. Bond}
\affil{Institute for Natural and Mathematical Sciences, Massey University, Private Bag 102904 North Shore Mail Centre, Auckland 0745, New Zealand}

\author[0000-0002-6578-5078]{E. Bachelet}
\affil{Las Cumbres Observatory, 6740 Cortona Drive, Suite 102, Goleta, CA 93117 USA}

\author[0000-0001-8014-6162]{D. Veras}
\affil{Centre for Exoplanets and Habitability, University of Warwick, Coventry CV4 7AL, UK}
\affil{Department of Physics, University of Warwick, Coventry CV4 7AL, UK}

\author[0000-0003-2302-9562]{N. Koshimoto}
\affil{Laboratory for Exoplanets and Stellar Astrophysics, NASA/Goddard Space Flight Center, Greenbelt, MD 20771, USA}
\affil{Department of Astronomy, Graduate School of Science, The University of Tokyo, 7-3-1 Hongo, Bunkyo-ku, Tokyo 113-0033, Japan}

\author[0000-0002-9782-0333]{V. Batista}
\affil{Sorbonne Universit\'es, UPMC Universit\'e Paris 6 et CNRS, UMR 7095, Institut d'Astrophysique de Paris, 98 bis bd Arago,
75014 Paris, France}

\author[0000-0002-7901-7213]{J.-B. Marquette}
\altaffiliation{Associated with Sorbonne Universit\'es, UPMC Universit\'e Paris 6 et CNRS, 
UMR 7095, Institut d'Astrophysique de Paris, 98 bis bd Arago,
75014 Paris, France}
\affil{Laboratoire d'astrophysique de Bordeaux, Univ. Bordeaux, CNRS, B18N, all\'ee Geoffroy Saint-Hilaire, 33615 Pessac, France}

%%ABSTRACT

\begin{abstract}
\textbf{Studies \citep{Vanderberg2015,Manser2019} have shown that remnants of destroyed planets and debris-disk planetesimals can survive the volatile evolution of their host stars into white dwarfs {\citep{Villaver2007,Duncan1998}}, but detection of intact planetary bodies around white dwarfs are few {\citep{Sigurdsson2003, Vanderburg2020, Luhman2011, Gaensicke2019}}. Simulations predict {\citep{Madappatt2016, Mustill2018, Norhaus2013}} that planets in Jupiter-like orbits around stars of {$\simlt 8 M_\odot$} avoid being destroyed by the strong tidal forces of their stellar host, but as yet there has been no observational confirmation of such a survivor. Here we report on the non-detection of a main-sequence lens star in the microlensing event MOA-2010-BLG-477Lb {\citep{Bachelet2012}} using near-infrared observations from the Keck Observatory. We determine this system contains a $0.53 \pm 0.11$ solar mass white dwarf host orbited by a $1.4 \pm 0.3$ Jupiter mass planet with a separation on the plane of the sky of $2.8 \pm 0.5$ AU, which implies a semi-major axis larger than this. This system is evidence that planets around white dwarfs can survive the giant and asymptotic giant phases of their host's evolution, and supports the prediction that over half of white dwarfs are predicted to have Jovian planetary companions {\citep{Schreiber2019}}. Located at approximately $2.0\,\si{kpc}$ toward the center of our Galaxy, it likely represents an analog to the end stages of the Sun and Jupiter in our own Solar System.}\\
\end{abstract}
\section*{} \label{sec:intro}
The microlensing event MOA-2010-BLG-477Lb was first detected by the Microlensing Observations in Astrophysics collaboration on 2 August 2010 {\citep{Bachelet2012}}. Microlensing is a technique which is sensitive to cold planets down to the mass of Earth {\citep{Bennett1996}} and can probe objects around all kinds of stars, including white dwarfs, as unlike other detection methods it does not rely on the light coming from the host. MOA-2010-BLG-477 is a planetary microlensing event with a planet-host mass ratio of $q = (2.61 \pm 0.03) \times 10^{-3}$ and a large Einstein ring radius  $\theta_E = 1.26 \pm  0.06$ mas, which implies a host star that is relatively massive or nearby.\\
\indent The target was subsequently observed with the \hl{Near Infra-Red Camera-2 (NIRC2)} instrument on the Keck-II telescope on July 27, 2015, August 5, 2016 and May 23, 2018 in the $H$ and $K_s$ near-infrared bands ($1.5-2.3\,\mu$m) using laser guide star adaptive optics. Data were obtained with the wide camera (40 arcsecond field-of-view) for calibration, and the narrow camera (10 arcsecond field-of-view) in an attempt to resolve the source and lens. Assuming it sits on the main sequence, we predict a host star with mass of $0.15 M_\odot < M_* < 0.93 M_\odot$, orbited by a planet with mass $0.5 M_{\rm Jup} < m_p < 2.1 M_{\rm Jup}$
at a distance of $0.7 < D_L < 2.7\,$kpc if all stellar types are equally likely to host a planet. 
Modelled parameters of the microlensing light curve, and the source magnitude and color {\citep{Boyajian2014}}, yield a lens-source relative proper motion of $\mu_{rel,G} = 11.53 \pm 0.56\,\si{mas/yr}$. This prediction, presented here in the geocentric reference frame that moves with the Earth's velocity at the time of the event, allows us to estimate the future lens-source separation following the event's peak on 12 August 2010.\\
\begin{figure*}[t]
\centering
%\vspace{3mm}
\includegraphics[width=1.00\textwidth]{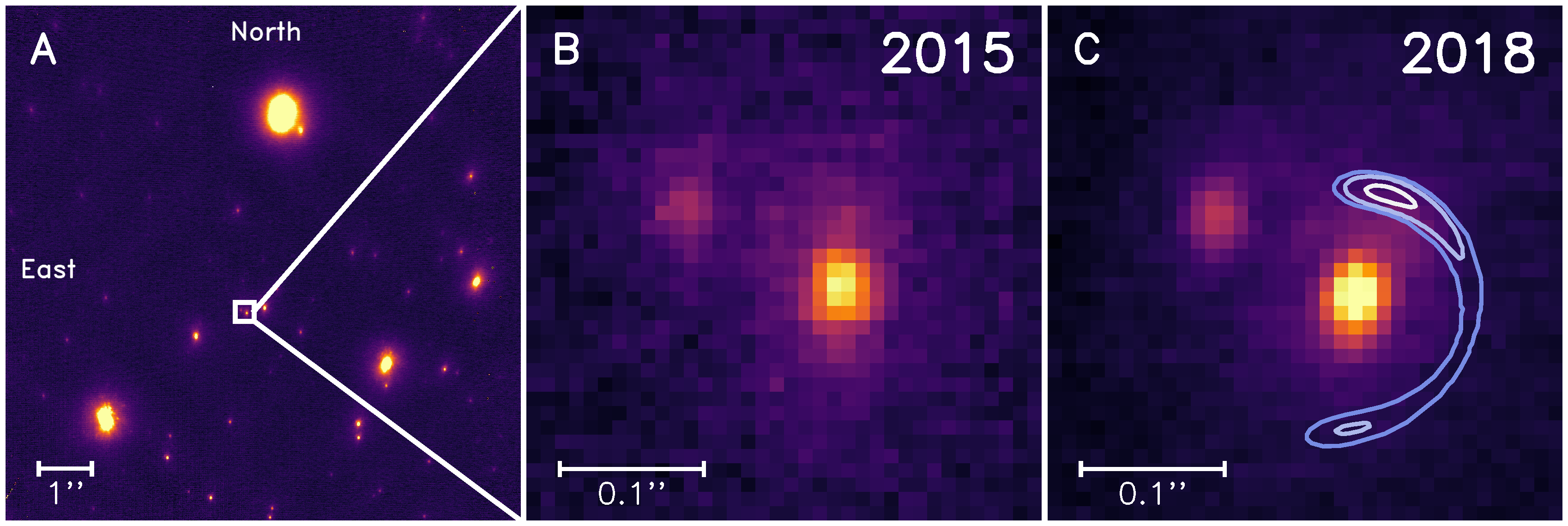}
\caption{\textbf{H-band adaptive optics imaging of MOA 2010-BLG-477 from the KECK observatory} \textbf{a.} A crop of a narrow-camera H-band image obtained with the NIRC2 imager in 2015 centered on MOA 2010-BLG-477 with an 8 arcsec field of view. \textbf{b.} A 0.36 arcsec zoom of the same image. The bright object in the center is the source. To the north-east (the upper left) is an unrelated $H=18.52\pm0.05$ star 123 mas from the source. \textbf{c.} The field in 2018. The contours indicate the likely positions of a possible main sequence host (probability of 0.393, 0.865, 0.989 from light to dark blue) using constraints from microlensing parallax and lens-source relative proper motion.}
\label{fig:keck}
\end{figure*}
\indent In our Keck images we find a $H=18.52\pm0.05$ star located ${\sim}123\,$mas to the north-east (upper-left in Fig.~\ref{fig:keck}c) of the source. Given the large lens-source relative proper motion, a predicted separation of $57\,\si{mas}$ means the lens star should be detectable under ideal observing conditions in 2015, 4.96 years after the event peak. By 2018, this separation will have widened to $90\,\si{mas}$. We find that the separation of the ${\sim}123\,$mas star to the north east has actually decreased by $11.5\,\si{mas}$ between 2015 and 2018, instead of the expected $33\,\si{mas}$ increase if it were the lens. The direction of this proper motion indicates that this object is unrelated to the lens and the planetary system.\\
\indent \hl{The source magnitude was measured as $H_s = 17.32 \pm 0.03$ and $K_s = 17.17 \pm 0.04$ using data} from the Cerro Tololo Inter-American Observatory (CTIO) and the Vista Variables in the Via Lactea (VVV) survey. This brightness is compatible with our NIRC2/KECK images, determined using the flux-ratio between the unrelated ${\sim}123\,$mas companion and the star at the position of the source. The sum of the two stars shown in Fig. {\ref{fig:keck}}c is $K=16.78\pm 0.02$ using the Keck data and $K=16.79\pm0.03$ using the CTIO data. This is consistent with there being no excess flux within the point spread function (PSF) of the source star. To confirm this, we model the PSF with the addition of weak contributions from the wings of the unrelated companion, and find no significant structure in the residuals. Hence, with no evidence of a detectable lens in our Keck/NIRC2 images, the lens star must have a brightness below the detection limit.\\
\indent To determine this limit we define a detection to be above the noise at the 3$\sigma$ level. This corresponds to a threshold of H ${\sim}21.1$, which means that any object brighter than this value should be detectable within our Keck narrow frames. \hl{This limit includes an extinction correction of $A_H=0.21$}. The light curve of this event indicates a microlensing parallax signal due to the orbital motion of the Earth, but it is not fully constrained. This implies that for each light curve model in our  Markov Chain distribution, we can determine the lens mass $M_L = (c^2/4G)(\theta_E/\pi_E)$ \citep{Gaudi2012}, where $\pi_E$ is the microlensing parallax. We combine these light curve constraints with limits on the source and lens distances and velocities, together with an empirical mass-luminosity relation \citep{Bennett2014a}, and find that a main-sequence lens star must have a brightness of $H < 18.10$ at 99.0\% confidence. This is largely due to the light curve constraint on the angular Einstein ring radius, $\theta_E$, derived from the angular radius of the source star and the Einstein ring crossing time.\\
\begin{figure*}[t!]
\centering
%\vspace{3mm}
\includegraphics[width=0.85\textwidth]{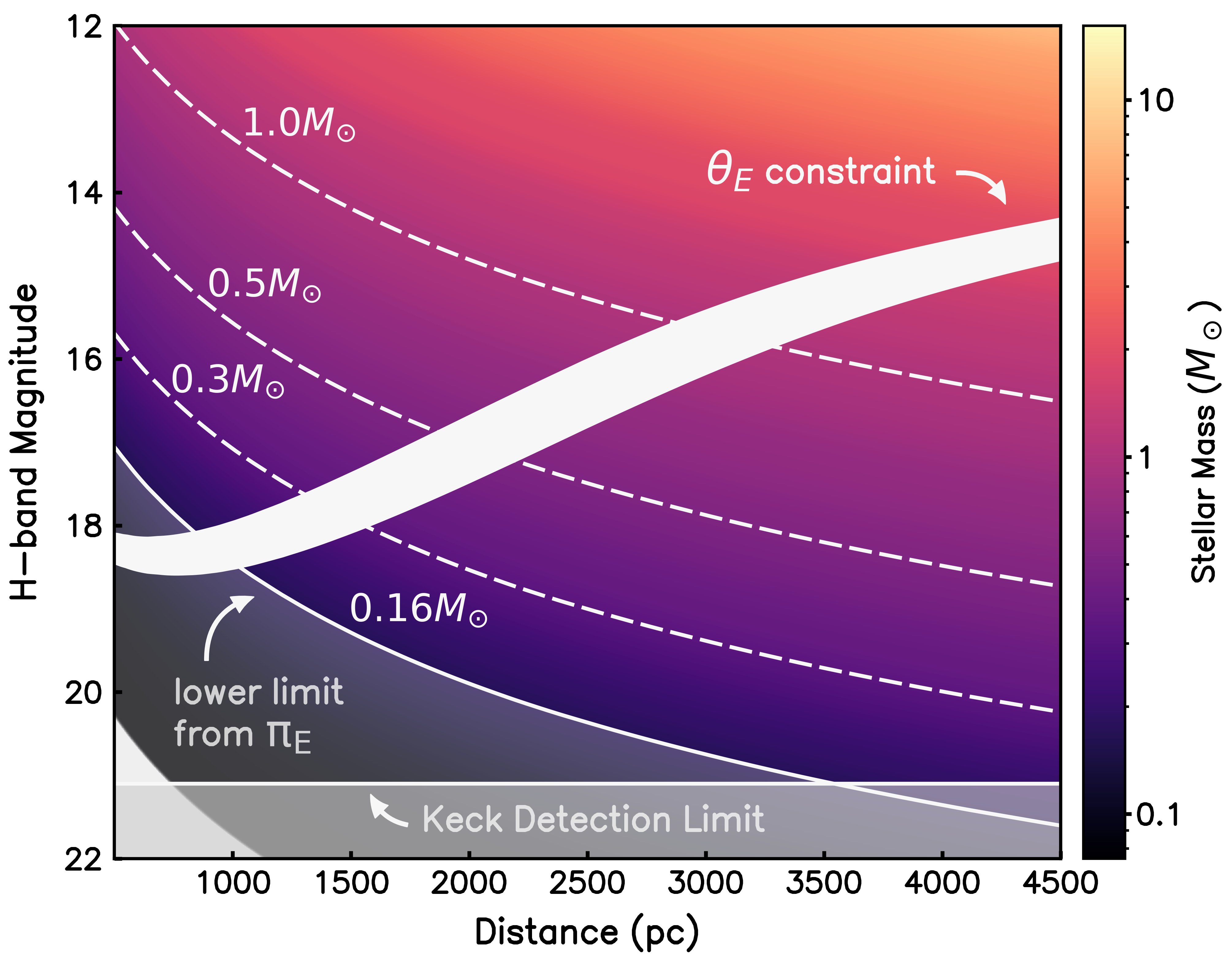}
\caption{\textbf{H-band brightness of possible main-sequence host lenses.} In white is the Einstein ring radius constraint, $\theta_E$, derived from finite source effects in the event light curve. This constraint indicates that if the planet host star was main-sequence, it should be visible with Keck adaptive optics as the entire area lies above the detection threshold of our H-band images at H${\sim}21.1$. The mass-luminosity relations for different main-sequence lens masses are shown: 1.0$M_\odot$, 0.5$M_\odot$, 0.3$M_\odot$, with 0.16$M_\odot$ the mass lower limit derived from the mircrolens parallax, $\pi_E$. Our null detection in our Keck images implies that the exoplanet host must be a stellar remnant, most likely a white dwarf.}
\label{fig:md}
\end{figure*}
\indent The microlensing parallax and lens-source relative proper motion measurements constrain the location of a main sequence lens star to the interior of the contours in Fig.~\ref{fig:keck}c. The predicted brightness of a main sequence lens as a function of lens distance can be see in Fig. \ref{fig:md}. Since all possible main-sequence lenses for the event are brighter than the Keck detection limit, and no such star is observed, the lens cannot be a main-sequence star. This same analysis also excludes brown dwarf lenses due to an upper limit on the microlensing parallax parameter, $\pi_E < 1.03$, which leads to an implied limit on the lens system mass of $M_L > 0.15 M_\odot$. Similarly, the lower microlensing parallax limit of $\pi_E > 0.26$, implies an upper mass limit of $M_L < 0.78 M_\odot$, which rules out neutron star and black hole host stars. With main sequence stars, brown dwarfs neutron stars, and black holes ruled out, we conclude that the lens must be a white dwarf.\\
\indent To estimate the properties of a white dwarf host, we use a complete sample of 130 white dwarfs within 20 pc \citep{Giammichele2012}, excluding unresolved double white dwarfs and double white dwarfs candidates identified by \cite{Toonen2017}, and the Galactic model from \citep{Bennett2014a}. This calculation was made under the assumption that all white dwarfs are equally likely to host planets. Our results are summarized in Fig. \ref{fig:Bayesian} and Table. \ref{tab:results}. We find a likely white dwarf host mass $M = 0.53 \pm 0.11$ M$_\odot$, which sits slightly below the peak of the single white dwarf mass distribution at 0.57-0.58 M$_\odot$ and excludes the high mass tail \citep{Tremblay2016}. This implies a Jovian planet of mass $m_p = 1.43\pm0.30$ M$_{J}$ at a distance of $D_L = 1.99 \pm 0.35 \,\si{kpc}$.\\
\begin{deluxetable*}{cccc}
\tablewidth{\textwidth}
\tablecaption{System Parameters of MOA 2010-BLG-477Lb\label{tab:results}}
\tablehead{\colhead{Parameter}& \colhead{Units}& \colhead{Value ($\pm1\sigma$)}& \colhead{2$\sigma$ range}}
\startdata
White Dwarf Lens Distance&$D_L$ (kpc)  & $1.99 \pm 0.35$ & $1.31$--$2.69$ \\
White Dwarf Lens Mass&$M_L\;(M_\odot)$  &  $0.53 \pm 0.11$ & $0.32$--$0.74$\\
Planet Mass&$m_p\;(M_J)$                & $1.43 \pm 0.30$ & $0.87$--$2.03$\\
Source Star Distance&$D_S$ (kpc)        & $7.8 \pm 1.3$ & $5.5$--$10.2$\\
2D star-planet separation&$a_\perp$ (AU) & $2.8 \pm 0.5$ & $1.9$--$3.7$\\  
3D star-planet separation&$a$ (AU) & $3.4^{+1.8}_{-0.8}$ & $2.1$--$12.1$\\  
Mass Ratio&$q\;(10^{-3})$   & $2.61 \pm 0.03$& $2.55$--$2.67$\\ 
White Dwarf Host $V_L$ &mag   & $26.92 \pm 1.80$ & $23.38$--$29.44$\\ 
White Dwarf Host $J_L$ &mag   & $24.93 \pm 0.91$ & $22.80$--$26.41$\\ 
\enddata
\end{deluxetable*}
\begin{figure*}[t]
\centering
%\vspace{3mm}
\includegraphics[width=0.9\textwidth]{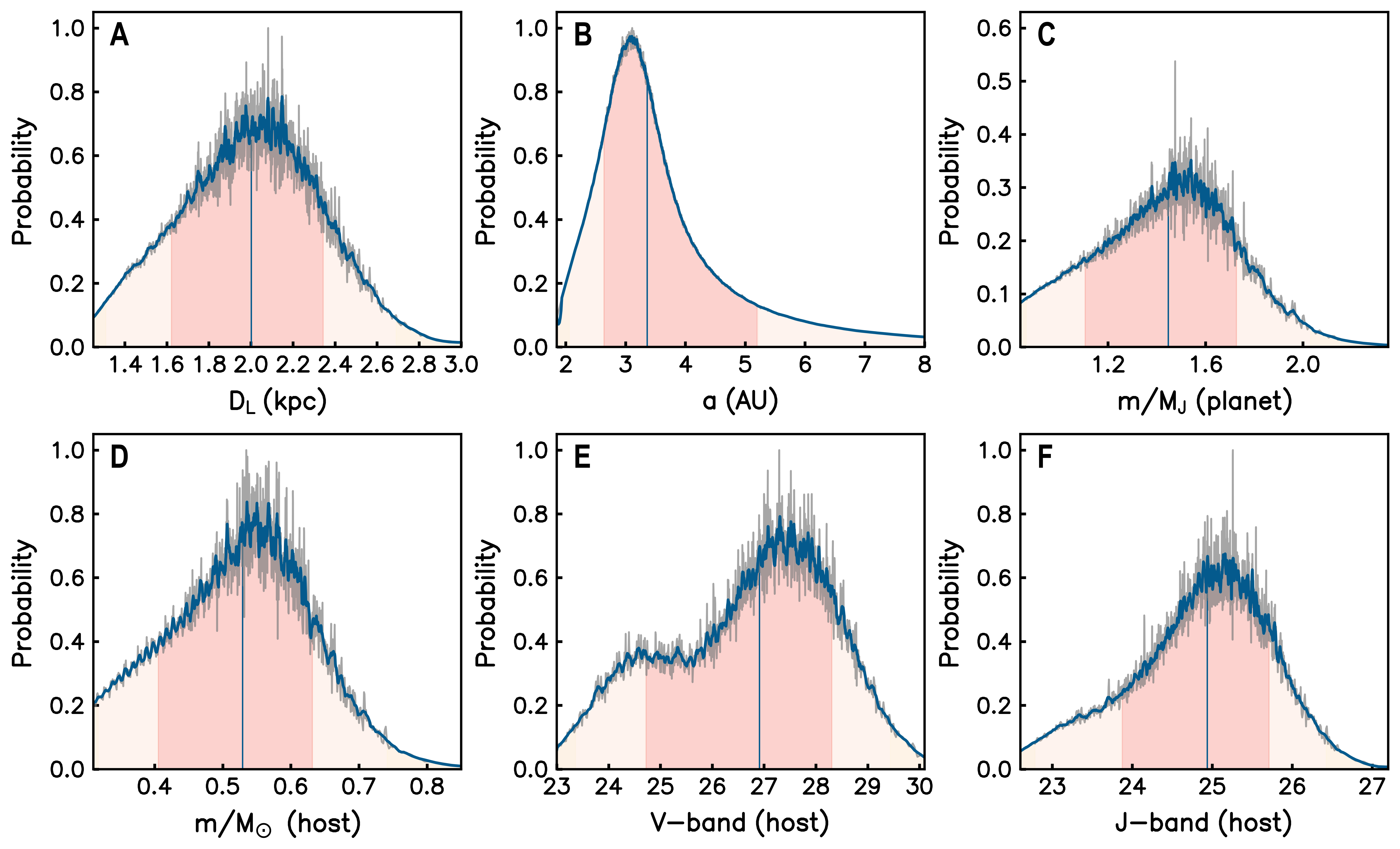}
\caption{\textbf{Physical properties of MOA-2010-BLG-477} Presented here are the (\textbf{a}) predicted distance to the host star/lens planetary system, (\textbf{b}) 3d star-planet separation, (\textbf{c}) mass of the planet, (\textbf{d}) mass of the host, (\textbf{e}) predicted host star brightness in V, and (\textbf{f}) predicted host star brightness in J. These parameters were calculated using a Bayesian analysis with priors from the relative proper motion of disk and bulge lenses, the lens star mass function, and velocity distributions derived from a galactic model as in {\cite{Bennett2014a}}}
\label{fig:Bayesian}
\end{figure*}
\indent With a separation from the white dwarf of $a = 3.4^{+1.8}_{-0.8}\,$AU (assuming a random orientation), it is likely that the planet MOA-2010-BLG-477Lb formed at the same time as the host star and managed to survive the post-main-sequence evolution. The mass loss experienced by a star on the giant and asymptotic giant branches pushes the planet toward a wider orbit, but tidal forces can have the opposite effect when the star expands to radii $\simgt 1\,$AU \citep{Veras2016}. In rare cases, the tidal effect can nearly cancel the mass loss effect, leaving a giant planet orbiting at a separation as small as $\sim 2\,$AU, but this requires a fine tuning of parameters to prevent the planet from being engulfed by the star. Most white dwarfs in the Galactic disk are thought to have formed from stars with initial masses of 1--$2.5\,M_\odot$ \citep{Veras2016}, and the measured separation means that to avoid tidal engulfment the mass of the progenitor star is likely to be $<2\,M_\odot$. Jovian planets orbiting these stars are generally thought to move to orbital separations $> 5$ or $6\,$AU around the remnant white dwarfs \citep{Mustill2012, Norhaus2013}. While this is larger than the $\sim2.8$ AU projected separation of MOA-2010-BLG-477Lb, in the region where microlensing has the highest sensitivity, the reduced detection probability for white dwarf planets might be compensated for by a higher intrinsic planet occurrence for gas giant planets around massive stars {\citep{Ghezzi2018}}. 

\indent While Mercury and Venus will most likely be engulfed by the Sun when it reaches the RGB phase, Jupiter is expected to survive {\citep{Veras2016}}. MOA-2010-BLG-477Lb is an example of the possible fate of Jupiter and supports the prediction that over half of white dwarfs are predicted to have Jovian planetary companions {\citep{Schreiber2019}}. It joins 29 other planets included in recent statistical sample of microlensing events {\citep{Suzuki2016}}. This will be supplemented with detections made by the Galactic Exoplanet Survey of the Nancy Grace Roman Space Telescope {\citep{Penny2019}} aimed at determining planet occurrence rates for white dwarf hosts. 

\section*{\textbf{Methods}} \label{sec:ao}
\subsection*{\normalfont \textbf{Observations}} \label{sec:ao}

\indent The microlensing event MOA 2010-BLG-477 was originally detected using the 1.8m telescope at Mt. John Observatory in New Zealand on August 2, 2010, and subsequently observed by more than 20 telescopes \citep{Bachelet2012}. In order to find the predicted magnitude of the source star we refer to the H-band light curve obtained by the $\mu$FUN 1.3 m SMARTS telescope at the Cerro Tololo Inter-American Observatory (CTIO). We calibrate the light curve according to data from the VVV (Vista Variables in the Via Lactea) survey \citep{Vandorou2020}. Using a single amplified CTIO H-band frame with an epoch corresponding to available VVV data, we cross identify between the two. We derive a source magnitude of $H_{CTIO,source} = 17.32 \pm 0.03$. Given the intrinsic source color of $(H-K)_0=0.07$, and using extinction corrections derived using the OGLE extinction calculator\footnote{OGLE Extinction Calculator,
\href{http://ogle.astrouw.edu.pl/cgi-ogle/getext.py}{http://ogle.astrouw.edu.pl/cgi-ogle/getext.py}} \citep{ Nataf2013}, $A_H=0.21$ and $A_K=0.13$, we find $K_{CTIO,source} = 17.17 \pm 0.04$.

Located at ($\alpha, \delta) = (18^h06^m07^s.47, -31^{\circ}27'16.17\arcsec$, J2000.0), we observed the event using the NIRC2 instrument on the Keck II telescope located on Mauna Kea (Hawaii) with laser guide star adaptive optics (LGSAO). Seven dithered H-band images were obtained with the narrow camera on 2015 July 27 (HJD = 2457230.7), 4.96 years following the event peak. These images have a median full width at half maximum (FWHM) of 50 mas. During this same epoch we obtained 14, 95 mas $K_s$-band images with the wide camera, and 10, 50 mas $H$-band images with the narrow camera. The event was observed twice more: in 2016 (HJD = 2457605.0) with 17 $H$-band 50 mas narrow images and two $H$-band 90 mas wide, and in 2018 with 16 50mas $K_s$-band narrow frames. %and 8 $H$-band narrow (both FWHM 50mas).
These final 2018 observations were taken on 23 May 2018 (HJD = 2458262.4), 7.78 years after maximum magnification.\\
\renewcommand{\figurename}{Extended Data Fig.}
\renewcommand\thefigure{\arabic{figure}} 
\setcounter{figure}{0}  
\begin{figure*}[h]
\centering
\includegraphics[width=0.85\textwidth]{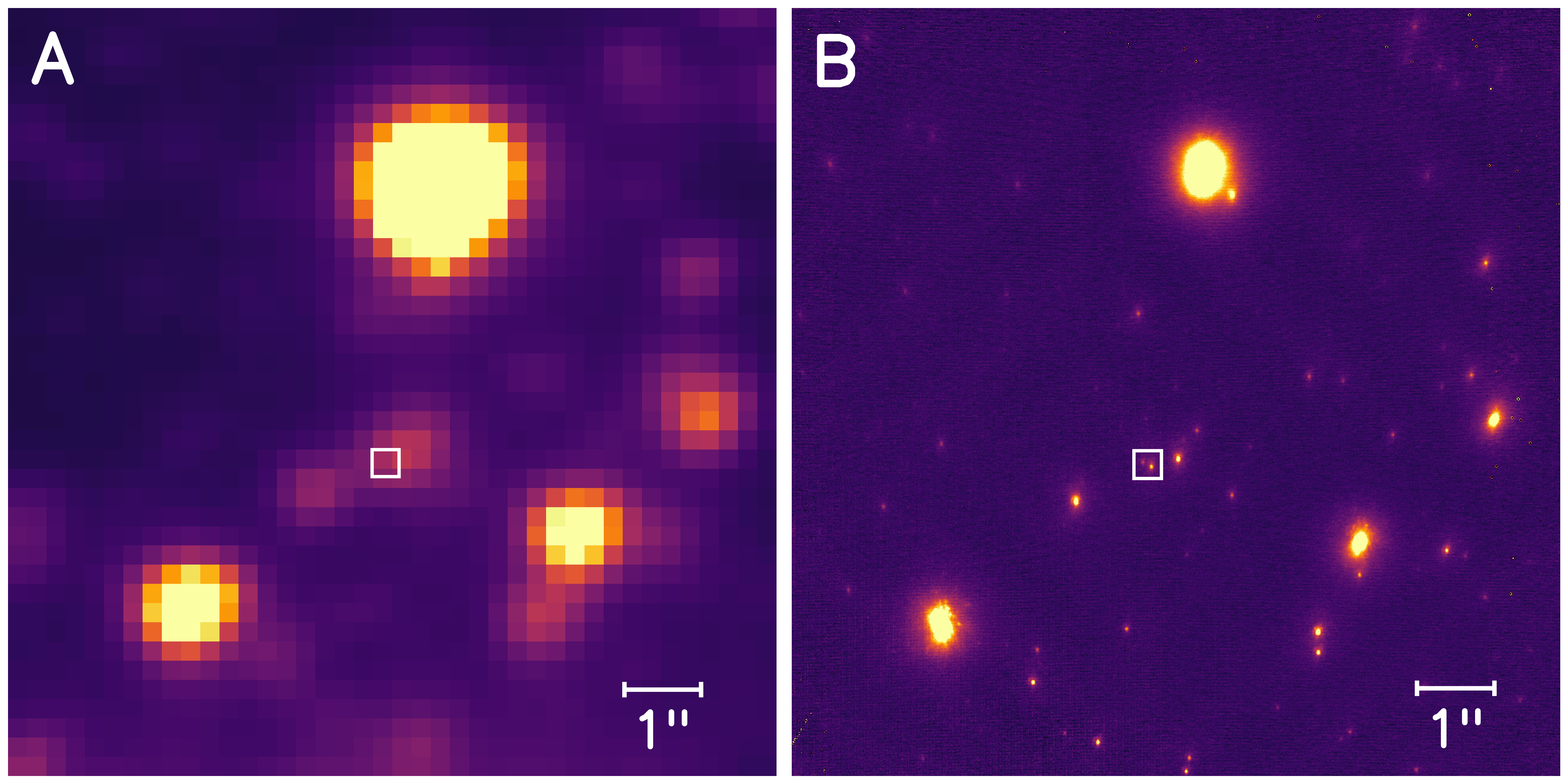}
\caption{\textbf{OGLE-III and Keck imaging of MOA-2010-BLG-477Lb.} (\textbf{a}) OGLE-III image of the OGLE-BLG176.8 field (\textbf{b}) H-band image of the same field taken in 2015 with Keck/NIRC2 with the narrow camera.}
\label{fig:ogle}
\end{figure*}
\hspace{-0.3cm} The goal of these observations was to determine the composition of the source-lens blend by obtaining a refined photometric and astrometric solution. We apply flat-field and dark-field corrections using standard techniques \citep{Blackman2020, Beaulieu2016}. The images were stacked using \texttt{SWARP} \citep{Bertin2010}. We identify the source+lens star as the bright object in Fig. {\ref{fig:keck}}c. Comparing this with the OGLE-III reference image of the BLG 176.8 field in Extended Data Fig. \ref{fig:ogle} we see that the OGLE star is a blend of four dimmer stars, the second brightest of which is the source+lens star. The OGLE star (number 119416) has an $I$-band magnitude of $I = 17.446 \pm 0.052$, which is consistent with the value of $I = 17.443 \pm 0.031$ from DoPhot CTIO photometry \citep{Bachelet2012}.\\
\indent We perform aperture photometry using \texttt{SExtractor} \citep{Bertin1996} on the wide K-band frame from 2015, using the narrow images to cross-calibrate the narrow frames, and find a K-band magnitude of the blend at the position of the source to be $K_{Blend}=16.78\pm0.03$. The magnitude is the combination of the flux from the source, the lens and the 123mas star located to the north-east. Fortunately, data of this field was also captured as part of the VVV survey while the event was still magnified. Data were obtained from the Vista telescope in JHK at the epochs:
\begin{equation}
\begin{aligned}
\mathrm{{H},MJD-OBS = 55423.1515}\\
\mathrm{{J},MJD-OBS = 55423.1576}\\
\mathrm{{K},MJD-OBS = 55423.1546}\\
\end{aligned}
\end{equation}\\
We use these data to determine whether the blend at the position of the source is consistent with the predicted source brightness. The image at the location of the source in the VVV image is a combination of the source, the lens, the $123\,$mas star to the north-east (upper left) and the star to the north-west (upper right). The brightness of these unrelated stars in the PSF are negligible when the source was amplified by 19.27 in H (where $\mathrm{MJD-OBS = 55423.15153784}$ and MJD = JD - 2400000.5). We compare the flux ratios between the star at the position of the source and the $123\,$mas companion, in the narrow Keck image and compare that to a calibrated wide Keck frame and that determined by CTIO/VVV.\\
\indent In our Keck wide image we crossmatch to stars in the VVV catalog \citep{Minniti2010}. Both the wide and narrow cameras on NIRC2 result in an image with dimensions of 1048 × 1048 pixels. The wide has a field of view (FOV) of 40 arcseconds while the narrow has a FOV of 10 arcseconds. The narrow images were taken in sequence with a dither of 0.7 arcseconds. The flux ratio of between the source star and the $123\,$mas companion is:
\begin{equation}
\begin{aligned}
\mathrm{{2018,K} = 0.42}\\
\mathrm{{2015,H} = 0.33}\\
\mathrm{{2016,H} = 0.34}\\
\end{aligned}
\end{equation}\\
Using the flux ratio from the 2018 K data and the predicted source K brightness of $K_s=17.17\pm0.04$, we determine the sum of the source and the $123\,$mas companion to be $K=16.79\pm0.04$, which is compatible with the the
$K_{Blend}=16.78\pm0.03$. determined from only the 2015 Keck wide co-added frame. This indicates that the there is no additional flux at the position of the source, and that all the photons from that object come from the source.\\

\subsection*{\normalfont \textbf{Light Curve Model}} \label{sec:model}
Since the publication of the MOA-2010-BLG-477Lb discovery paper {\citep{Bachelet2012}}, the MOA group has developed detrending methods \citep{Bennett2012,Bond2017} that can remove systematic errors due to the color dependence of atmospheric refraction. We expect a significant color dependent differential refraction signal due to a bright main sequence star at $I = 13.5$ that is only 3 arc seconds from the MOA-2010-BLG-477 source star, and we have found a significant correction using the method of Bond {\citep{Bond2017}}. The best fit model parameters from using this new data
are shown in Extended Data Table~{\ref{tab-mparams}}. Three of the parameters also apply to single lens models. These are the Einstein radius crossing time, $t_E$, the time of closest approach, $t_0$ between the source and the lens center-of-mass, and the distance of this closest approach, $u_0$, in units of the Einstein radius. A binary lens requires three additional parameters, the lens separation, $s$, in units of the Einstein radius, the angle between the lens separation and source trajectory, $\alpha$, and the planet-star mass ratio, $q$. Events like MOA-2010-BLG-477 with caustic crossings require the source radius crossing time, $t_*$, to account for finite source effects. Finally, it is also important to include parameters to describe the orbital motion of the lens system and the Earth-based observers. The orbital period of the planet is many years, so the only detectable effect of orbital motion is the relative lens velocity, $\dot{s}_x$ and $\dot{s}_y$, in the directions parallel and perpendicular to the lens separation. The microlensing parallax is described by the North and East components of the microlensing parallax vector, $\piEbold$. The inclusion of microlensing parallax introduces an approximate degeneracy that corresponds to a flip in the orientation of the lens system with respect to the orientation of the Earth's orbit. We label the solutions with the sign of the $u_0$ parameter. The $u_0 > 0$ model is favored by $\Delta\chi^2 = 5.08$, and it is displayed in Extended Data Fig. {\ref{fig:pi_relWD}}, but both models are compatible with our overall conclusions.\\
\begin{figure*}[t]
\centering
%\vspace{3mm}
\includegraphics[width=0.8\textwidth]{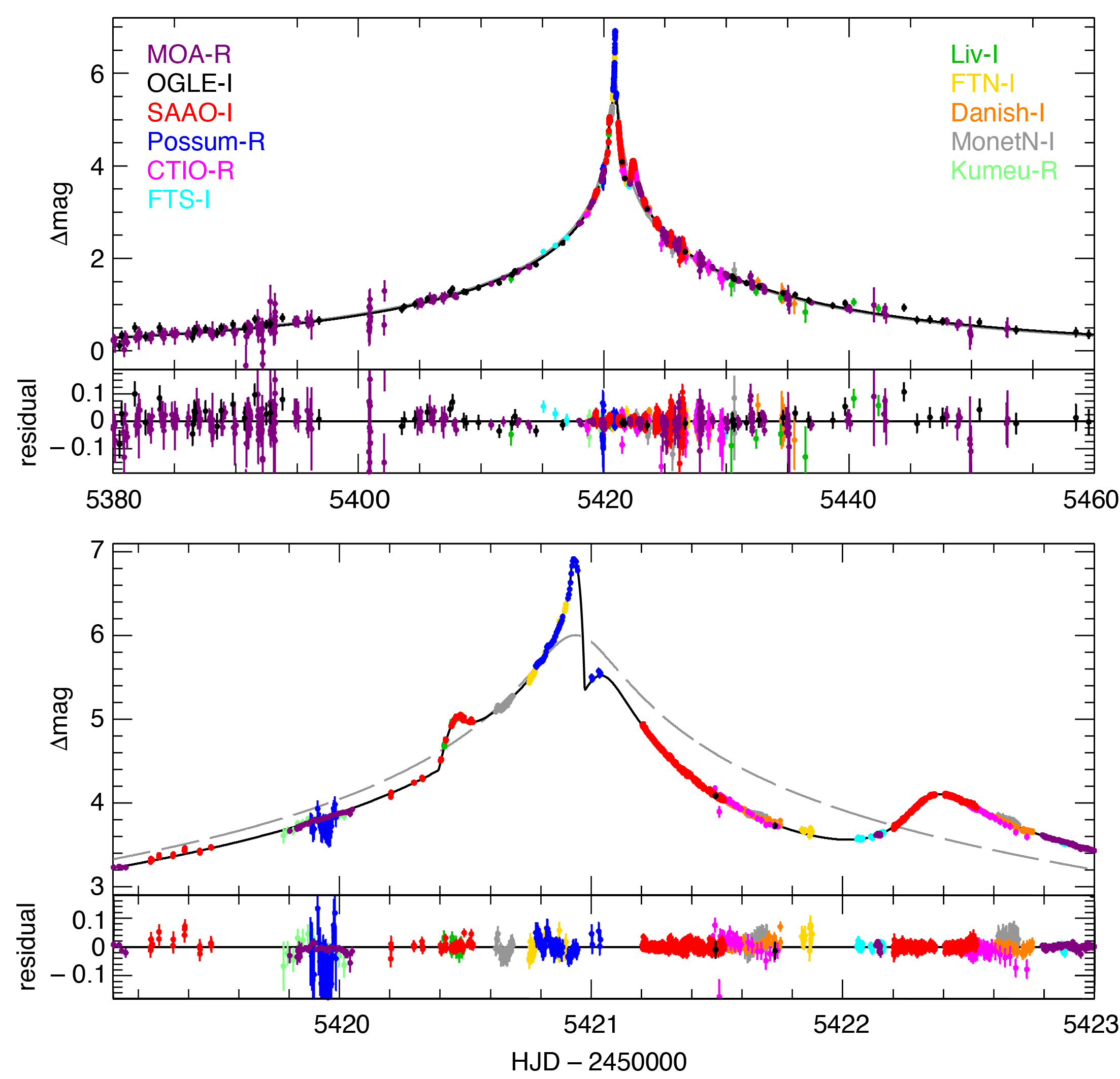}
\caption{\textbf{Light Curve Data and Model} for microlensing event MOA-2010-BLG-477. The solid curve is the best fit model and the dashed grey curve is the single lens model with the same single lens parameters. The different colors represent different data sets from different telescopes. \hl{One sigma error bars are shown.} The data sets are explained in the discovery paper {\citep{Bachelet2012}}.}
\label{fig:lc}
\end{figure*}
\renewcommand{\tablename}{Extended Data Table.}
\setcounter{table}{0} 
\begin{deluxetable}{ccccc}
\tablecaption{Best Fit Model Parameters
                         \label{tab-mparams} }
\tablewidth{0pt}
\tablehead{
%% Use a footnote to explain numbering.
 & \multicolumn{2}{c} {This Analysis} & \multicolumn{2}{c} {Discovery Paper} \\
\colhead{parameter}  & \colhead{$u_0 > 0$} & \colhead{$u_0 < 0$}  & \colhead{$u_0 > 0$} & \colhead{$u_0 < 0$}  
}  % end header.
\startdata
$t_E$ (days) & $39.523(425)$  & $39.796(441)$ & $46.868$ & $47.560$ \\   
$t_0$ (${\rm HJD}^\prime$) & $5420.9381(3)$  & $5420.9376(3)$ & $5420.9382$ & $5420.9374$ \\
$u_0$ & $0.003966(51)$ & $-0.003937(54)$ & $0.003430$ & $-0.003315$\\
$s$ & $1.12203(52)$ & $1.12278(51)$ & $1.12407$ & $1.12429$\\
$\alpha$ (rad) & $0.54126(176)$ & $-0.54137(174)$ & $0.54004$ & $-0.53836$ \\
$q \times 10^{3}$ & $2.590(31)$ & $2.587(31)$ & $2.210$ & $2.168$ \\
$t_\ast$ (days) & $0.02736(18)$ & $0.02739(18)$ & $0.02719$ & $0.02719$\\
$\dot{s}_x$ & $0.00039^{+0.00021}_{-0.00030}$ & $0.00030^{+0.00023}_{-0.00027}$ & $0.00223$ & $0.00016$\\
$\dot{s}_y$ & $-0.00192^{+0.00099}_{-0.00146}$ & $0.00274^{+0.00042}_{-0.00194}$ & $-0.00587$ & $0.00196$ \\
$\pi_{\rm E,N}$ & $0.4450^{+0.0750}_{-0.1499}$ & $-0.5393^{+0.4053}_{-0.1524}$ & $0.5432$ & $0.0974$\\
$\pi_{\rm E,E}$ & $-0.3165^{+0.0723}_{-0.0250}$ & $-0.1467^{+0.0420}_{-0.0766}$ & $-0.1026$ & $-0.0134$ \\
$\chi^2/{\rm dof}$ & $6420.91/6490$ & $6425.97/6490$ & - & - \\
\enddata
\tablecomments{All errors are given as $\pm1\sigma$.}
\end{deluxetable}
\hspace{-0.3cm} One major difference between the analysis in the discovery paper {\citep{Bachelet2012}} and our new analysis is that the Einstein ring crossing time, $t_E$, has dropped by 
about 16\%. This causes a similar increase in the mass ratio, $q$, because the timescale of the planetary signal has not changed. The other significant change in the model parameters can be found in the microlensing parallax values. As shown in Extended Data Fig. {\ref{fig:pi_relWD}}a, only one component of the microlensing parallax vector, $\piEbold$, is tightly constrained by the data. It is quite common for the component parallel to the direction of the Earth's acceleration at the time of the event to be measured much more precisely than the perpendicular component, and in this case the uncertainty in the perpendicular component is increased by the degeneracy between microlensing parallax and lens orbital motion. This parallel component is roughly in the the East-West direction for events in the Galactic bulge. However, the full $\piEbold$ can be determined \citep{Bhattacharya2018,Bennett2020} with a measurement of the lens-source relative proper motion, $\boldsymbol{\mu}_\mathrm{rel,G}$, although this must be be done in the Geocentric frame that has been used for the light curve modeling.\\ %This is discussed in more detail in Section~{\ref{sec:detect}}.\\
\indent While the changes in the model due to the improved MOA data are quite noticeable, they have very little effect on our conclusions. We have performed the full analysis using the light curve models from the discovery paper {\citep{Bachelet2012}}, and these implied very similar results. The only difference is that the old analysis shifted the $\pi_{E,E}$ toward more positive values by 0.1--0.2. As a result, the old light curve analysis allowed higher mass host stars.\\
\indent We note that the discovery paper suggested that it might be sensible to search for three body lens systems that could explain this light curve. However, unlike the case of published triple lens events \citep{Gaudi2008,Bennett2010a,Gould2014,Bennett2016a}, this light curve shows no evidence of an unmodeled light curve feature that can be explained by an additional lens. Thus, we can expect that the only effect of an additional lens would be to increase the uncertainties in the properties of the lens masses in the binary lens model {\citep{Zhu2014}}, usually by very small amounts. Since our primary conclusions are based on the properties of the host star, an additional lens that is consistent with the data would not affect our conclusions.

\subsection*{\normalfont \textbf{Bayesian Analysis Light Curve Models with a Galactic Prior}} \label{sec:detect}

Our analysis makes use of the light curve models from the Markov Chain Monte Carlo calculations from our reanalysis of the MOA data with our detrending method {\citep{Bond2017}}, as well as other data sets from the discovery paper {\citep{Bachelet2012}}. These models incorporate the effects of the orbital motion of the Earth which are responsible for the microlensing parallax, as well as the orbital motion of the planet. The microlensing parallax is a two-dimensional vector, $\piEbold$, parallel to the lens-source relative proper motion, which means the distribution of microlensing parallax vectors from the light curve models constrains the direction of the lens-source relative proper motion. The amplitude of the lens-source relative proper motion vector is determined by finite source effects in the microlensing light curve. However, the direction and length of
this lens-source relative proper motion vectors are not determined in the Heliocentric reference frame that is appropriate for high angular resolution Keck follow-up observations. The light curve models employ a
Geocentric reference frame that moves with the velocity of the Earth at the time of closest lens-source
alignment. For MOA-2010-BLG-477, this velocity is $\bm{v}_{\Earth N,E}=(-2.7933, 19.5634)\: \mathrm{km\: s^{-1}}$,
and the transformation to the relative proper motion in the Heliocentric reference frame, 
$\boldsymbol{\mu}_\mathrm{rel,H}$, is given by
\begin{equation} \label{eq:helio} 
\begin{aligned}
\boldsymbol{\mu}_\mathrm{rel,H} = \boldsymbol{\mu}_\mathrm{rel,G} + \frac{\boldsymbol{v}_{\oplus\mathrm{N,E}}}{\mathrm{AU}} \pi_\mathrm{rel} \ ,
\end{aligned}
\end{equation}
\citep{Dong2007,Bhattacharya2018},
where the relative parallax is given by $\pi_{\rm rel}=\mathrm{AU}/D_L-\mathrm{AU}/D_S$.
This transformation from $\boldsymbol{\mu}_\mathrm{rel,G}$ to $\boldsymbol{\mu}_\mathrm{rel,H}$ cannot be
computed with light curve parameters only as $\pi_\mathrm{rel}$ depends on the source and lens 
distances. Therefore, we must invoke a Galactic model to properly sample the source distance, $D_S$, values. We have used the Galactic model from \cite{Bennett2014a} in our analysis. Our results are largely independent of the choice of Galactic model for this Bayesian analysis. The only poorly understood prior is the host mass dependence of the planet hosting probability, since we are considering a lens that is known to host a planet. For simplicity, we assume that all white dwarfs are equally likely to host a planet.  Once the source distance 
is selected, the lens distance can be determined from light curve parameters, 
\begin{equation}
D_L = {{\rm AU}\over \pi_{\rm E}\theta_{\rm E}+{\rm AU}/D_S} \ .
\label{eq-Dl}
\end{equation}
\begin{figure*}[t!]
\centering
%\vspace{3mm}
\includegraphics[width=0.85\textwidth]{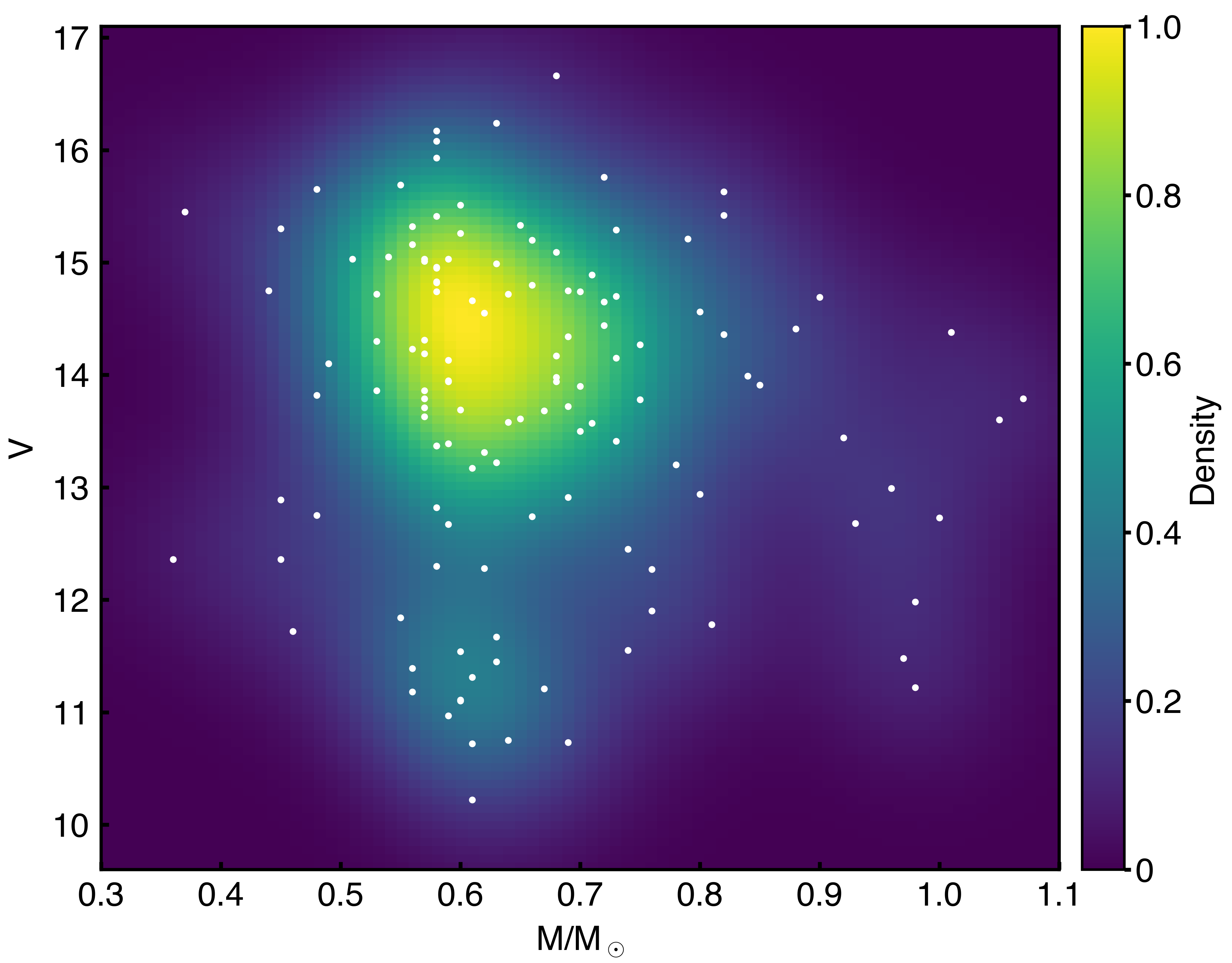}
\caption{\textbf{White Dwarf Mass-Luminosity distribution} derived a sample of 130 white dwarfs from a homogeneous and complete sample of white dwarfs within 20pc of the Sun \cite{Giammichele2012}. Two unresolved double-white dwarfs (DWD), eight unresolved DWD candidates and one unresolved binary white dwarf with a main-sequence companion have been removed from this sample \citep{Toonen2017}. 14 stars with distances $> 20\,$pc have also been removed. The white dots indicate the masses and $V$ band magnitudes of the white dwarfs in this sample, and the color distribution indicates the smooth Gaussian multivariate kernel-density distribution that we have used in our analysis.}
\label{fig:Bayesian}
\end{figure*}

\begin{figure*}[t!]
\centering
%\vspace{3mm}
\includegraphics[width=0.7\textwidth]{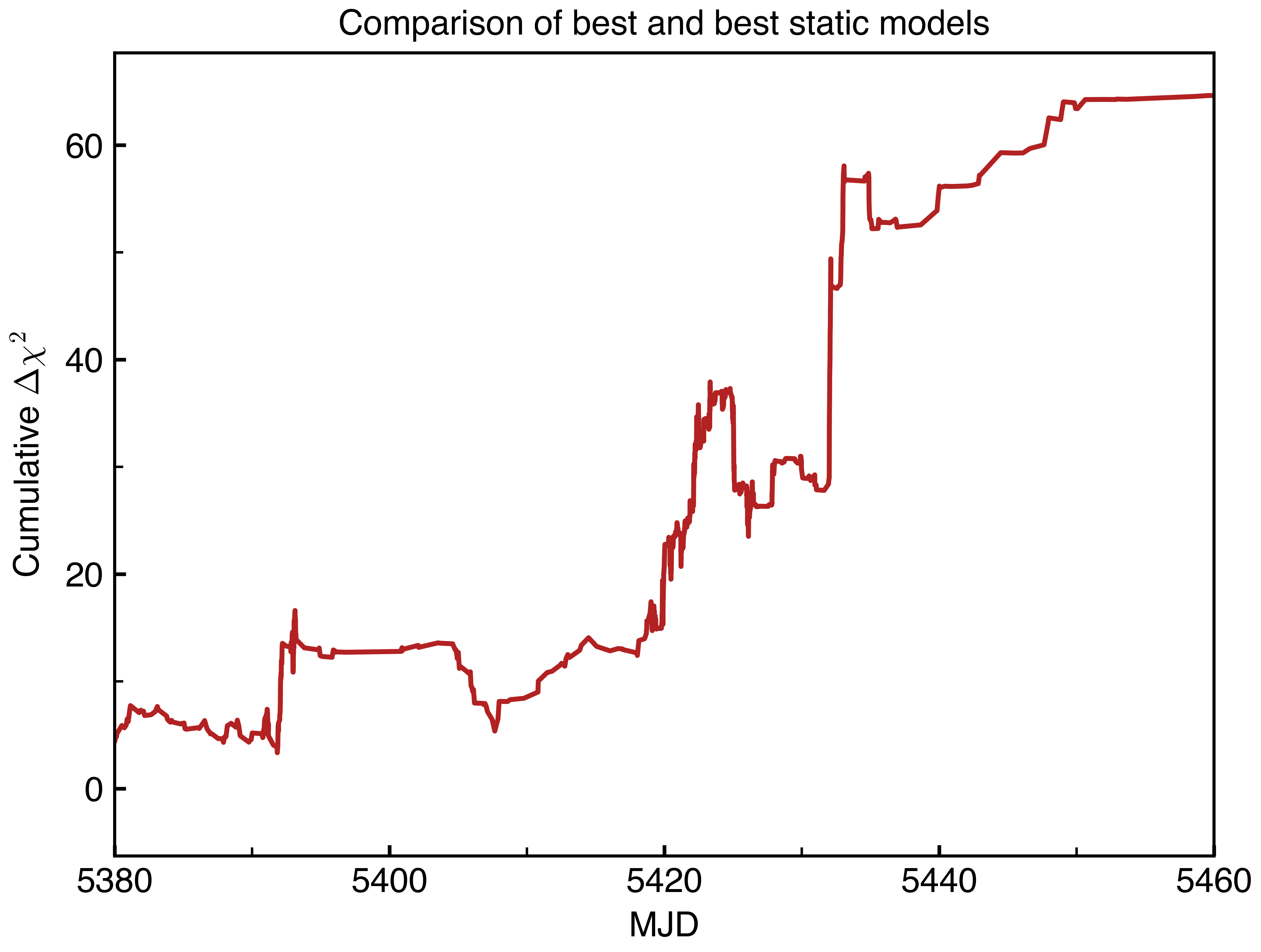}
\caption{\textbf{Cumulative $\Delta\chi^2$ comparing a static light curve model with that including parallax and orbital motion.} The bulk of the signal comes following the light curve peak. The parallax plus orbital motion comes primarily from the MOA data ($\Delta\chi^2=45$) and the SAAO data ($\Delta\chi^2=9.0$).}
\label{fig:cumulativedist}
\end{figure*}
\begin{figure*}[h]
\centering
%\vspace{3mm}
\includegraphics[width=0.8\textwidth]{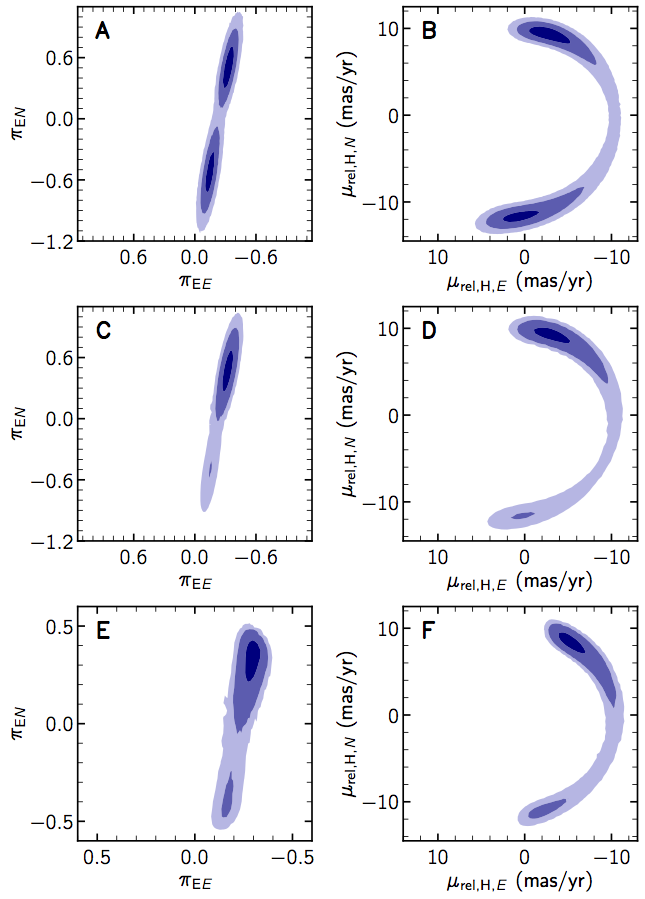}
\caption{\textbf{Predictions of the microlens parallax vector $\bm{\pi}_E$ and the corresponding predicted relative lens-source proper motion $\bm{\mu}_{rel}$} for a main sequence and white dwarf lens. Based on a Markov-Chain Monte-Carlo (MCMC) analysis using Galactic model priors as in {\cite{Bennett2014a}}, the upper panels (\textbf{a}) and (\textbf{b}) show the unweighted predicted components of $(\pi_{EN},\pi_{EE})$ and $(\mu_{rel,HN},\mu_{rel,HE})$. The middle panels (\textbf{c}) and (\textbf{d}) show the weighted predictions for a main-sequence lens. The lower panels (\textbf{e}) and (\textbf{f}) show the weighted predictions for a white-dwarf lens. The three shades of blue from dark to light denote probabilities of of 0.393, 0.865, 0.989. When integrating over all parameters the limit of the 0.393 contour corresponds to the 1$\sigma$ distribution of any chosen parameter.}
\label{fig:pi_relWD}
\end{figure*}
\begin{figure*}[h]
\centering
%\vspace{3mm}
\includegraphics[width=0.9\textwidth]{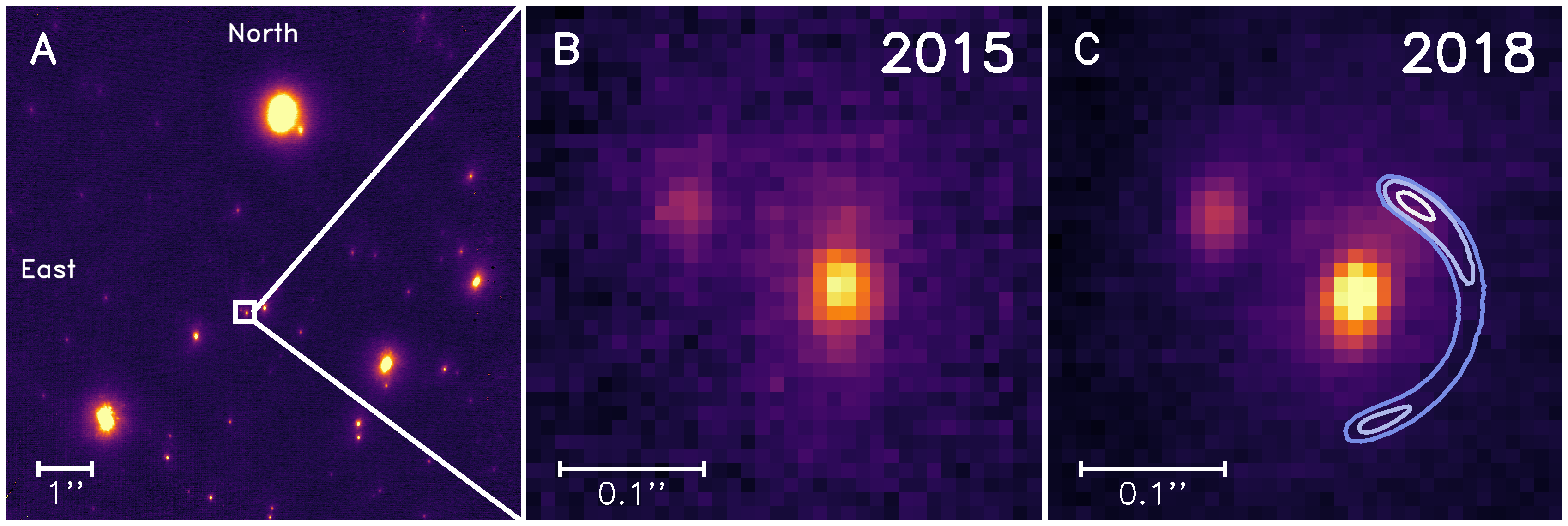}
\caption{\textbf{H-band adaptive optics imaging from the KECK observatory, with contours showing the predicted position of a white dwarf lens (analogous to Fig. {\ref{fig:keck}})} (\textbf{a}) A crop of a narrow-camera H-band image obtained with the NIRC2 imager in 2015 centered on MOA 2010-BLG-477 with an 8 arcsec field of view. (\textbf{b}) A 0.36 arcsec zoom of the same image.The bright object in the center is the source. To the north-east (the upper left) is an unrelated $H= 18.52\pm0.05$ star $123\,$mas from the source (star 123NE. (\textbf{c}) The field in 2018. The contours indicate the likely positions of the white dwarf host (probability of 0.393, 0.865, 0.989 from light to dark blue) using constraints from microlensing parallax and lens-source relative proper motion.}
\label{fig:pi_relWD_Keck}
\end{figure*}
While $\pi_{\rm E}$ is only partially constrained by the light curve data, it is precisely determined for each model in the Markov Chain, so equation~{\ref{eq-Dl}} can be used for each Markov Chain model.
The Galactic model also provides weights for the density of stars at the distance to the lens, $D_L$, and by the probability of a star with a mass equal to the lens mass. This implicitly includes the assumption that all stars are equally likely to host the planet of the measured mass ratio and projected separation.\\
Extended Data Fig.~\ref{fig:pi_relWD} shows how the Galactic model affects the distributions of the $\boldsymbol{\pi}_\mathrm{E}$ and $\boldsymbol{\mu}_\mathrm{rel,H}$ vectors. Panels \ref{fig:pi_relWD}a and \ref{fig:pi_relWD}b show the $\boldsymbol{\pi}_\mathrm{E}$ distribution from the light curve models in \citep{Bachelet2012} and the implied $\boldsymbol{\mu}_\mathrm{rel,H}$ distribution, with the help of the $D_S$ distribution from the \citep{Bennett2014a} Galactic model. The component of the $\boldsymbol{\pi}_\mathrm{E}$ parallel to the direction of the Earth's acceleration during the event is tightly constrained. This component is close to the East direction. Conversely the perpendicular component, which is largely in the North-South direction is very weakly constrained. % and the distribution of $\boldsymbol{\pi}_\mathrm{E}$ values runs through the origin, where the magnitude of the microlensing parallax vector $\pi_\mathrm{E} = 0$, which would correspond to an infinite lens mass.
Panels c and d of Extended Data Fig.~\ref{fig:pi_relWD} show the result when we apply the complete Galactic model including the mass function for main sequence stars. This removes the light curve models with small $\pi_\mathrm{E}$ values, and therefore, large masses. It was these low-$\pi_\mathrm{E}$, high-mass 
light curve models that allowed the $\boldsymbol{\pi}_\mathrm{E}$ vector to point in any direction. This was responsible for the ring distribution of $\boldsymbol{\mu}_\mathrm{rel,H}$, but with the high mass lens
systems excluded, this ring is broken into two arcs to the North and South, with low-mass lens systems only allowed
in the Northern arc. Our Keck observations have ruled out any main sequence stellar lenses in these arcs.
Our model does assume that stars with masses
above $1.1\,M_\odot$ have left the main sequence, but such stars would also be brighter than the source star,
which are clearly ruled out over the full ring in Extended Data Fig. \ref{fig:pi_relWD}b ring.

With main sequence hosts ruled out, we can now turn our attention to white dwarf host stars for the
MOA-2010-BLG-477Lb exoplanet. We can repeat the same calculation that we did for main sequence sources.
This requires a mass-luminosity distribution for white dwarfs. We construct such a distribution from the 
20\,pc sample as in \citep{Giammichele2012}, excluding unresolved binary white dwarfs because these are likely
to have unreliable parameters. The resulting white dwarf mass-luminosity relation, constructed using a multivariate Gaussian kernel density-estimation is shown in Extended Data Fig.~\ref{fig:Bayesian}.
The results of a repeat of the Bayesian analysis with the main sequence mass function replaced by our 
white dwarf mass function, from Extended Data Fig.~\ref{fig:Bayesian}, is shown in Extended Data Fig.~\ref{fig:pi_relWD} panels e and f. The results are 
quite similar to the main sequence case, but the exclusion of low-$\pi_\mathrm{E}$, high-mass 
light curve models is now somewhat stronger because the white dwarf mass function is strongly peaked around $M_L \sim 0.57-0.58 M_\odot$. This Bayesian analysis is also used to produce the lens system properties 
presented in Table~\ref{tab:results}. The contours from Extended Data Fig.~\ref{fig:pi_relWD}d are 
reproduced in Extended Data Fig.~\ref{fig:pi_relWD_Keck}c, where they replace the contours for the main sequence
stars shown in Fig.~\ref{fig:keck}c.

\subsection*{\normalfont \textbf{Detection Limits and Point Spread Function Deconvolution}}

We determine the detection threshold of our Keck images by estimating the flux of a point source and evaluating the normalized cross-product with the point-spread function (PSF). We obtain this quantity for all points in the subtracted image and construct a map of the amplitudes. In order to obtain an estimate of the fluctuations due to the noise we calculate the standard deviation in this map. We consider that these fluctuations are significant at the 3$\sigma$ level and that this 3$\sigma$ level is our detection limit. We then convert this 3$\sigma$ limit to magnitude and obtain a minimum limit of detection of $H~\simeq$ 21.1.\par
\begin{figure*}[h]
\centering
\includegraphics[width=0.85\textwidth]{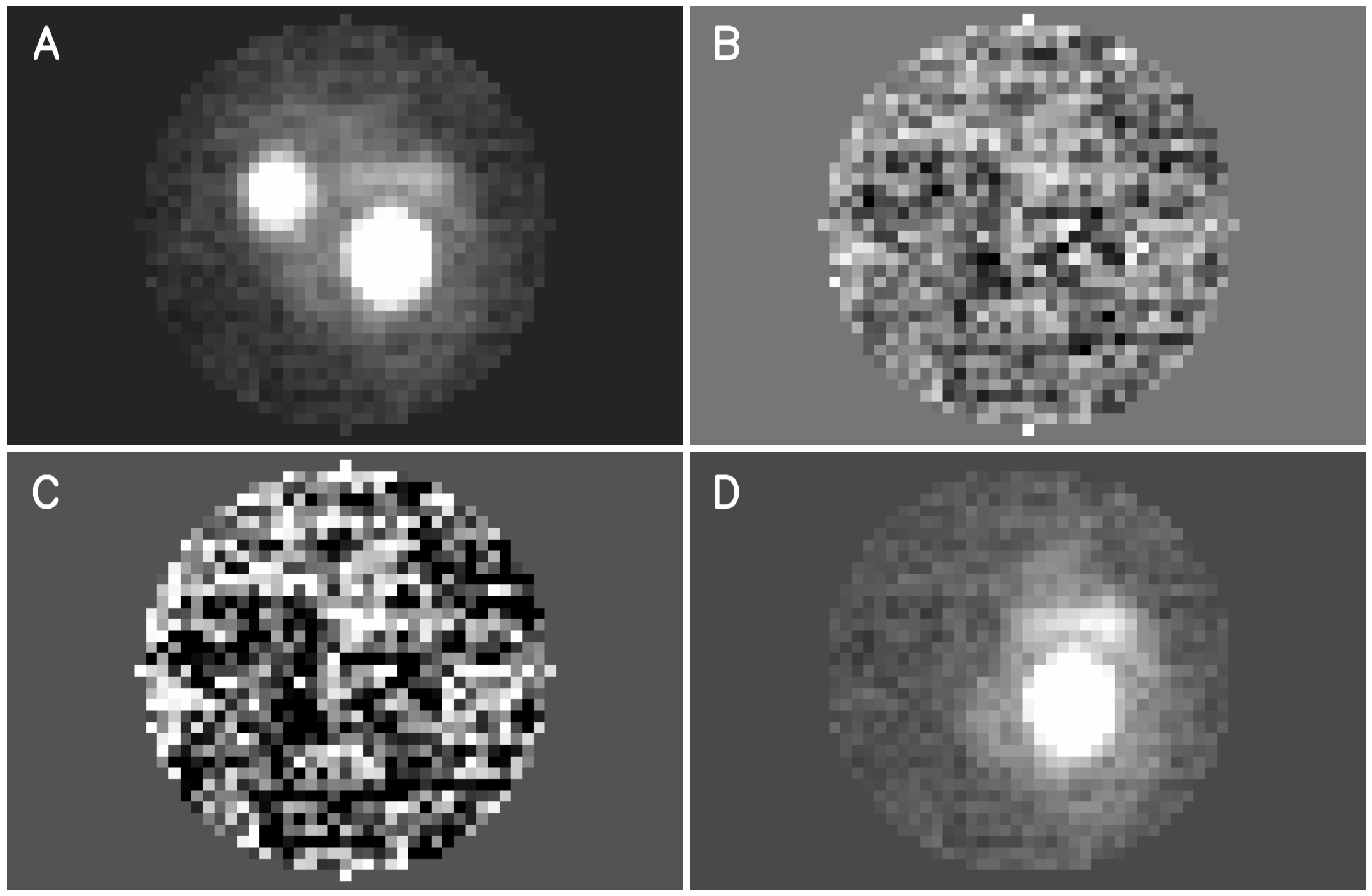}
\caption{\textbf{Keck Point Spread Function (PSF) fit and residuals.} (\textbf{a}) Keck/NIRC2 H-band image from 2018. (\textbf{b}) Residuals after fitting the PSF using multiple stars in the neighbourhood of the source. Both the object to the north-east (upper left in panel A) and the object at source position (lower right) are subtracted using this PSF fit. There is no structure or indication of a double star in either of the two objects. (\textbf{c}) The residuals from panel B normalized to the Poisson noise. (\textbf{d}) Panel A but subtracting the fitted PSF from the unrelated companion.}
\label{fig:psf}
\end{figure*}

While the relative proper motion indicates that the lens is distinct from the PSF of the source star, we then perform an analysis to determine if the star at the position of the source is consistent with a single object or if there is evidence of a two component system. To do this we make a numerical estimate of the PSF by stacking the brightest stars in the neighborhood of our target. The accurate position of each PSF is estimated by iterative Gaussian weighted centering. The PSFs are then interpolated, re-centered and stacked, and median solution obtained. In our H-band images the star of interest is quite close to the star to the north-east, which means it may receive some contribution from the PSF wings of its neighbour. To subtract any neighbour contributions we reconstruct the wings of the PSF of the more distant star as a single function of distance and subtract this weak contribution from the image. Then the best solution obtained is subtracted, leaving only the stars of interest. The result of this subtraction and the residuals are presented in Extended Data Fig. \ref{fig:psf}.\par
When reconstructing a single star a common problem is to determine whether this star is a single PSF or a very close system of multiple PSFs. If the single PSF fit results in significant residuals, for example, it is clear that the fit of a binary system should be attempted. In this case constraints on the PSF can be used to find upper limits on allowed separations that are compatible with the data and noise. Our method to achieve this and to overcome the degeneracy created when the PSF components are very close are described below.\par
The image as shown in Extended Data Fig. \ref{fig:psf}, $I_C$ is the result of the convolution of the object data, $I_0$ with the PSF, $\phi$. Let's consider the case where the un-convolved data $I_0$ is very narrow with respect to the PSF, which is typical in cases of two close-together stars. 
In this instance, we can write:
\begin{equation}
I_{C}(x, y)=\int \phi(x-u, y-v) I_{0}(u, v) d u d v
\end{equation}

Considering that the variations are small with respect to the variations of
\(\phi,\) we can write \(\phi\) as a local expansion in the local variables \((u, v)\):
\begin{equation}
I_{C}(x, y) \simeq m_{0} \phi(x, y)-m_{1} \frac{\partial \phi}{\partial x}-m_{2} \frac{\partial \phi}{\partial y}+m_{3} \frac{\partial^{2} \phi}{\partial x^{2}}+m_{4} \frac{\partial^{2} \phi}{\partial x \partial y}+m_{5} \frac{\partial^{2} \phi}{\partial y^{2}}
\end{equation}
where:\\
\begin{equation}
\begin{cases}
m_{0}=\int I_{0}(u, v) d u d v\\
m_{1}=\int I_{0}(u, v) u\; d u d v\\
m_{2}=\int I_{0}(u, v) v\; d u d v\\
m_{3}=\frac{1}{2} \int I_{0}(u, v) u^{2} d u d v\\
m_{4}=\int I_{0}(u, v) uv\; d u d v\\
m_{5}=\frac{1}{2} \int I_{0}(u, v) v^{2} d u d v\\
\end{cases}
\end{equation}
\\
The $m_1$ and $m_2$ coefficients represent the degrees of freedom related to the centered of the function $I_0$. To eliminate these degrees of freedom we make the center of flux of $I_0$ coincide with the origin of the coordinate system. In this case, $m_1=0$ and $m_2=0$. As such we are left with an expansion with 4 basis functions, the PSF and its 3 second order derivatives. The moments of the function \(I_{0}\) are simply the 3 coefficients \(\left(m_{3}, m_{4}, m_{5}\right)\) normalized by the total flux (Note that provided that the PSF is normalized the total flux and the coefficient \(m_{0}\) should be very similar).\par
We use the numerical model of the PSF to reconstruct the derivatives up to the second order. The derivatives are obtained by shifting the PSF model and taking the difference with the original PSF. The value of the shift is small with respect to the size of the PSF grid. We choose a value of 0.01, but also tested 0.1 and 0.001 and no significant changes were observed. In creating the PSF model we take an area around the object large enough to include the PSF wings but small enough to avoid including another object. 

\bibliographystyle{yahapj}
\bibliography{library}

\begin{thebibliography}{}
\providecommand\natexlab[1]{#1}
\providecommand\JournalTitle[1]{#1}

\bibitem[{Bachelet {et~al.}(2012)Bachelet, Shin, Han, Fouqu{\'{e}}, Gould,
  Menzies, Beaulieu, Bennett, Bond, Dong, Heyrovsk{\'{y}}, Marquette, Marshall,
  Skowron, Street, Sumi, Udalski, Abe, Agabi, Albrow, Allen, Bertin, Bos,
  Bramich, Chavez, Christie, Cole, Crouzet, Dieters, Dominik, Drummond,
  Greenhill, Guillot, Henderson, Hessman, Horne, Hundertmark, Johnson,
  J{\o}rgensen, Kandori, Liebig, M{\'{e}}karnia, McCormick, Moorhouse,
  Nagayama, Nataf, Natusch, Nishiyama, Rivet, Sahu, Shvartzvald, Thornley,
  Tomczak, Tsapras, Yee, Batista, Bennett, Brillant, Caldwell, Cassan,
  Corrales, Coutures, {Dominis Prester}, Donatowicz, Kubas, Martin, Williams,
  Zub, {Andrade de Almeida}, DePoy, Gaudi, Hung, Jablonski, Kaspi, Klein, Lee,
  Lee, Koo, Maoz, Mu{\~{n}}oz, Pogge, Polishook, Shporer, Abe, Botzler, Chote,
  Freeman, Fukui, Furusawa, Harris, Itow, Kobara, Ling, Masuda, Matsubara,
  Miyake, Ohmori, Ohnishi, Rattenbury, Saito, Sullivan, Suzuki, Sweatman,
  Tristram, Wada, Yock, Szyma{\'{n}}ski, Soszy{\'{n}}ski, Kubiak, Poleski,
  Ulaczyk, Pietrzy{\'{n}}ski, Wyrzykowski, Kains, Snodgrass, Steele, Alsubai,
  Bozza, Browne, Burgdorf, {Calchi Novati}, Dodds, Dreizler, Finet, Gerner,
  Hardis, Harps{\o}e, Hinse, Kerins, Mancini, Mathiasen, Penny, Proft, Rahvar,
  Ricci, Scarpetta, Sch{\"{a}}fer, Sch{\"{o}}nebeck, Southworth, Surdej, \&
  Wambsganss}]{Bachelet2012}
Bachelet, E., Shin, I.-G., Han, C., {et~al.} 2012,
  \href{http://dx.doi.org/10.1088/0004-637X/754/1/73}{\JournalTitle{Astrophys.
  J}, 754, 73}

\bibitem[{Beaulieu {et~al.}(2016)Beaulieu, Bennett, Batista, Fukui, Marquette,
  Brillant, Cole, Rogers, Sumi, Abe, Bhattacharya, Koshimoto, Suzuki, Tristram,
  \& Han}]{Beaulieu2016}
Beaulieu, J.-P., Bennett, D.~P., Batista, V., {et~al.} 2016,
  \href{http://dx.doi.org/10.3847/0004-637X/824/2/83}{\JournalTitle{Astrophys.
  J}, 824, 83}

\bibitem[{Bennett \& Rhie(1996)}]{Bennett1996}
Bennett, D.~P., \& Rhie, S.~H. 1996,
  \href{http://iopscience.iop.org/article/10.1086/178096/pdf}{\JournalTitle{Astrophys.
  J}, 472, 660}

\bibitem[{Bennett {et~al.}(2010)Bennett, Rhie, Nikolaev, Gaudi, Udalski, Gould,
  Christie, Maoz, Dong, McCormick, Szyma{\'{n}}ski, Tristram, MacIntosh, Cook,
  Kubiak, Pietrzy{\'{n}}ski, Soszy{\'{n}}ski, Szewczyk, Ulaczyk, Wyrzykowski,
  Depoy, Han, Kaspi, Lee, Mallia, Natusch, Park, Pogge, Polishook, Abe, Bond,
  Botzler, Fukui, Hearnshaw, Itow, Kamiya, Korpela, Kilmartin, Lin, Ling,
  Masuda, Matsubara, Motomura, Muraki, Nakamura, Okumura, Ohnishi, Perrott,
  Rattenbury, Sako, Saito, Sato, Skuljan, Sullivan, Sumi, Sweatman, Yock,
  Albrow, Allan, Beaulieu, Bramich, Burgdorf, Coutures, Dominik, Dieters,
  Fouqu{\'{e}}, Greenhill, Horne, Snodgrass, Steele, Tsapras, Chaboyer,
  Crocker, \& Frank}]{Bennett2010a}
Bennett, D.~P., Rhie, S.~H., Nikolaev, S., {et~al.} 2010,
  \href{http://dx.doi.org/10.1088/0004-637X/713/2/837}{\JournalTitle{Astrophys.
  J}, 713, 837}

\bibitem[{Bennett {et~al.}(2012)Bennett, Sumi, Bond, Kamiya, Abe, Botzler,
  Fukui, Furusawa, Itow, Korpela, Kilmartin, Ling, Masuda, Matsubara, Miyake,
  Muraki, Ohnishi, Rattenbury, Saito, Sullivan, Suzuki, Sweatman, Tristram,
  Wada, \& Yock}]{Bennett2012}
Bennett, D.~P., Sumi, T., Bond, I.~A., {et~al.} 2012,
  \href{http://dx.doi.org/10.1088/0004-637X/757/2/119}{\JournalTitle{Astrophys.
  J}, 757, 119}

\bibitem[{Bennett {et~al.}(2014)Bennett, Batista, Bond, Bennett, Suzuki,
  Beaulieu, Udalski, Donatowicz, Bozza, Abe, Botzler, Freeman, Fukunaga, Fukui,
  Itow, Koshimoto, Ling, Masuda, Matsubara, Muraki, Namba, Ohnishi, Rattenbury,
  Saito, Sullivan, Sumi, Sweatman, Tristram, Tsurumi, Wada, Yock, Albrow,
  Bachelet, Brillant, Caldwell, Cassan, \& Cole}]{Bennett2014a}
Bennett, D.~P., Batista, V., Bond, I.~A., {et~al.} 2014,
  \href{http://dx.doi.org/10.1088/0004-637X/785/2/155}{\JournalTitle{Astrophys.
  J}, 785, 155}

\bibitem[{Bennett {et~al.}(2016)Bennett, Rhie, Udalski, Gould, Tsapras, Kubas,
  Bond, Greenhill, Cassan, Rattenbury, Boyajian, Luhn, Penny, Anderson, Abe,
  Bhattacharya, Botzler, Donachie, Freeman, Fukui, Hirao, Itow, Koshimoto, Li,
  Ling, Masuda, Matsubara, Muraki, Nagakane, Ohnishi, Oyokawa, Perrott, Saito,
  Sharan, Sullivan, Sumi, Suzuki, Tristram, Yonehara, Yock, Szyma{\'{n}}ski,
  Soszy{\'{n}}ski, Ulaczyk, Wyrzykowski, Allen, DePoy, Gal-Yam, Gaudi, Han,
  Monard, Ofek, Pogge, Street, Bramich, Dominik, Horne, Snodgrass, Steele,
  Albrow, Bachelet, Batista, Beaulieu, Brillant, Caldwell, Cole, Coutures,
  Dieters, Prester, Donatowicz, Fouqu{\'{e}}, Hundertmark, J{\o}rgensen, Kains,
  Kane, Marquette, Menzies, Pollard, Ranc, Sahu, Wambsganss, Williams, \&
  Zub}]{Bennett2016a}
Bennett, D.~P., Rhie, S.~H., Udalski, A., {et~al.} 2016,
  \href{http://dx.doi.org/10.3847/0004-6256/152/5/125}{\JournalTitle{Astrophys.
  J}, 152, 125}

\bibitem[{Bennett {et~al.}(2020)Bennett, Bhattacharya, Beaulieu, Blackman,
  Vandorou, Terry, Cole, Henderson, Koshimoto, Lu, {Baptiste Marquette}, Ranc,
  \& Udalski}]{Bennett2020}
Bennett, D.~P., Bhattacharya, A., Beaulieu, J.-P., {et~al.} 2020,
  \href{http://dx.doi.org/10.3847/1538-3881/ab6212}{\JournalTitle{Astrophys.
  J}, 159, 68}

\bibitem[{Bertin \& Arnouts(1996)}]{Bertin1996}
Bertin, E., \& Arnouts, S. 1996,
  \href{http://dx.doi.org/10.1051/aas:1996164}{\JournalTitle{A{\&}AS}, 117,
  393}

\bibitem[{Bertin \& Emmanuel(2010)}]{Bertin2010}
Bertin, E., \& Emmanuel. 2010,
  \href{http://adsabs.harvard.edu/abs/2010ascl.soft10068B}{\JournalTitle{Astrophysics
  Source Code Library, ascl:1010.068}}

\bibitem[{Bhattacharya {et~al.}(2018)Bhattacharya, Beaulieu, Bennett, Anderson,
  Koshimoto, Lu, Batista, Blackman, Bond, Fukui, Henderson, Hirao, Marquette,
  Mroz, Ranc, \& Udalski}]{Bhattacharya2018}
Bhattacharya, A., Beaulieu, J.-P., Bennett, D.~P., {et~al.} 2018,
  \href{http://dx.doi.org/10.3847/1538-3881/aaed46}{\JournalTitle{Astrophys.
  J}, 156, 289}

\bibitem[{Blackman {et~al.}(2020)Blackman, Beaulieu, Cole, Vandorou, Koshimoto,
  Bachelet, Batista, Bhattacharya, \& Bennett}]{Blackman2020}
Blackman, J.~W., Beaulieu, J.-P., Cole, A.~A., {et~al.} 2020,
  \href{http://dx.doi.org/10.3847/1538-4357/ab68da}{\JournalTitle{Astrophys.
  J}, 890, 87}

\bibitem[{Bond {et~al.}(2017)Bond, Bennett, Sumi, Udalski, Suzuki, Rattenbury,
  Bozza, Koshimoto, Abe, Asakura, Barry, Bhattacharya, Donachie, Evans, Fukui,
  Hirao, Itow, Li, Ling, Masuda, Matsubara, Muraki, Nagakane, Ohnishi, Ranc,
  Saito, Sharan, Sullivan, Tristram, Yamada, Yamada, Yonehara, Skowron,
  Szyma{\'{n}}ski, Poleski, Mr{\'{o}}z, Soszy{\'{n}}ski, Pietrukowicz,
  Kozlowski, Ulaczyk, \& Pawlak}]{Bond2017}
Bond, I.~A., Bennett, D.~P., Sumi, T., {et~al.} 2017,
  \href{http://dx.doi.org/10.1093/mnras/stx1049}{\JournalTitle{Mon. Not. R.
  Astron. Soc.}, 469, 2434}

\bibitem[{Boyajian {et~al.}(2014)Boyajian, {Von Braun}, {Van Belle},
  Farrington, Schaefer, Jones, White, McAlister, {Ten Brummelaar}, Ridgway,
  Gies, Sturmann, Sturmann, Turner, Goldfinger, \& Vargas}]{Boyajian2014}
Boyajian, T.~S., {Von Braun}, K., {Van Belle}, G., {et~al.} 2014,
  \href{http://dx.doi.org/10.1088/0004-637X/787/1/92}{\JournalTitle{Astrophys.
  J}, 787, 92}

\bibitem[{Dong {et~al.}(2007)Dong, Udalski, Gould, Reach, Christie, Boden,
  Bennett, Fazio, Griest, Szyma{\'{n}}ski, Kubiak, Soszy{\'{n}}ski,
  Pietrzy{\'{n}}ski, Szewczyk, Wyrzykowski, Ulaczyk, Wieckowski,
  Paczy{\'{n}}ski, DePoy, Pogge, Preston, Thompson, \& Patten}]{Dong2007}
Dong, S., Udalski, A., Gould, A., {et~al.} 2007,
  \href{http://dx.doi.org/10.1086/518536}{\JournalTitle{Astrophys. J}, 664,
  862}

\bibitem[{Duncan \& Lissauer(1998)}]{Duncan1998}
Duncan, M.~J., \& Lissauer, J.~J. 1998,
  \href{http://dx.doi.org/10.1006/icar.1998.5962}{\JournalTitle{Icarus}, 134,
  303}

\bibitem[{Gaensicke {et~al.}(2019)Gaensicke, Schreiber, Toloza, Fusillo,
  Koester, \& Manser}]{Gaensicke2019}
Gaensicke, B.~T., Schreiber, M.~R., Toloza, O., {et~al.} 2019,
  \href{http://dx.doi.org/10.1038/s41586-019-1789-8}{\JournalTitle{Nature},
  576, 61}

\bibitem[{Gaudi(2012)}]{Gaudi2012}
Gaudi, B.~S. 2012,
  \href{http://dx.doi.org/10.1146/annurev-astro-081811-125518}{\JournalTitle{ARAAAJ},
  50, 411}

\bibitem[{Gaudi {et~al.}(2008)Gaudi, Bennett, Udalski, Gould, Christie, Maoz,
  Dong, McCormick, Szymanski, Tristram, Nikolaev, Paczynski, Kubiak,
  Pietrzynski, Soszynski, Szewczyk, Ulaczyk, Wyrzykowski, DePoy, Han, Kaspi,
  Lee, Mallia, Natusch, Pogge, Park, Abe, Bond, Botzler, Fukui, Hearnshaw,
  Itow, Kamiya, Korpela, Kilmartin, Lin, Masuda, Matsubara, Motomura, Muraki,
  Nakamura, Okumura, Ohnishi, Rattenbury, Sako, Saito, Sato, Skuljan, Sullivan,
  Sumi, Sweatman, Yock, Albrow, Allan, Beaulieu, Burgdorf, Cook, Coutures,
  Dominik, Dieters, Fouque, Greenhill, Horne, Steele, Tsapras, Chaboyer,
  Crocker, Frank, \& Macintosh}]{Gaudi2008}
Gaudi, B.~S., Bennett, D.~P., Udalski, A., {et~al.} 2008,
  \href{http://dx.doi.org/10.1126/science.1151947}{\JournalTitle{Science}, 319,
  927}

\bibitem[{Ghezzi {et~al.}(2018)Ghezzi, Montet, \& Johnson}]{Ghezzi2018}
Ghezzi, L., Montet, B.~T., \& Johnson, J.~A. 2018,
  \href{http://dx.doi.org/10.3847/1538-4357/aac37c}{\JournalTitle{Astrophys.
  J}, 860, 109}

\bibitem[{Giammichele {et~al.}(2012)Giammichele, Bergeron, \&
  Dufour}]{Giammichele2012}
Giammichele, N., Bergeron, P., \& Dufour, P. 2012,
  \href{http://dx.doi.org/10.1088/0067-0049/199/2/29}{\JournalTitle{ApJS}, 199,
  29}

\bibitem[{Gould {et~al.}(2014)Gould, Udalski, Shin, Porritt, Skowron, Han, Yee,
  Koz{\l}owski, Choi, Poleski, Wyrzykowski, Ulaczyk, Pietrukowicz, Mr{\'{o}}z,
  Szyma{\'{n}}ski, Kubiak, Soszy{\'{n}}ski, Pietrzy{\'{n}}ski, Gaudi, Christie,
  Drummond, McCormick, Natusch, Ngan, Tan, Albrow, DePoy, Hwang, Jung, Lee,
  Park, Pogge, Abe, Bennett, Bond, Botzler, Freeman, Fukui, Fukunaga, Itow,
  Koshimoto, Larsen, Ling, Masuda, Matsubara, Muraki, Namba, Ohnishi, Philpott,
  Rattenbury, Saito, Sullivan, Sumi, Suzuki, Tristram, Tsurumi, Wada, Yamai,
  Yock, Yonehara, Shvartzvald, Maoz, Kaspi, \& Friedmann}]{Gould2014}
Gould, A., Udalski, A., Shin, I.~G., {et~al.} 2014,
  \href{http://dx.doi.org/10.1126/science.1251527}{\JournalTitle{Science}, 345,
  46}

\bibitem[{Luhman {et~al.}(2011)Luhman, Burgasser, \& Bochanski}]{Luhman2011}
Luhman, K.~L., Burgasser, A.~J., \& Bochanski, J.~J. 2011,
  \href{http://dx.doi.org/10.1088/2041-8205/730/1/L9}{\JournalTitle{ApJL}, 730,
  9}

\bibitem[{Madappatt {et~al.}(2016)Madappatt, {De Marco}, \&
  Villaver}]{Madappatt2016}
Madappatt, N., {De Marco}, O., \& Villaver, E. 2016,
  \href{http://dx.doi.org/10.1093/mnras/stw2025}{\JournalTitle{Mon. Not. R.
  Astron. Soc.}, 463, 1040}

\bibitem[{Manser {et~al.}(2019)Manser, G{\"{a}}nsicke, Eggl, Hollands,
  Izquierdo, Koester, Landstreet, Lyra, Marsh, Meru, Mustill,
  Rodr{\'{i}}guez-Gil, Toloza, Veras, Wilson, Burleigh, Davies, Farihi,
  {Gentile Fusillo}, de~Martino, Parsons, Quirrenbach, Raddi, Reffert, {Del
  Santo}, Schreiber, Silvotti, Toonen, Villaver, Wyatt, Xu, \& {Portegies
  Zwart}}]{Manser2019}
Manser, C.~J., G{\"{a}}nsicke, B.~T., Eggl, S., {et~al.} 2019,
  \href{http://dx.doi.org/10.1126/science.aat5330}{\JournalTitle{Science}, 364,
  66}

\bibitem[{Minniti {et~al.}(2010)Minniti, Lucas, Emerson, Saito, Hempel,
  Pietrukowicz, Ahumada, Alonso, Alonso-Garcia, Arias, Bandyopadhyay,
  Barb{\'{a}}, Barbuy, Bedin, Bica, Borissova, Bronfman, Carraro, Catelan,
  Clari{\'{a}}, Cross, de~Grijs, D{\'{e}}k{\'{a}}ny, Drew, Fari{\~{n}}a,
  Feinstein, Laj{\'{u}}s, Gamen, Geisler, Gieren, Goldman, Gonzalez, Gunthardt,
  Gurovich, Hambly, Irwin, Ivanov, Jord{\'{a}}n, Kerins, Kinemuchi, Kurtev,
  L{\'{o}}pez-Corredoira, Maccarone, Masetti, Merlo, Messineo, Mirabel, Monaco,
  Morelli, Padilla, Palma, Parisi, Pignata, Rejkuba, Roman-Lopes, Sale,
  Schreiber, Schr{\"{o}}der, Smith, Jr., Soto, Tamura, Tappert, Thompson,
  Toledo, Zoccali, \& Pietrzynski}]{Minniti2010}
Minniti, D., Lucas, P.~W., Emerson, J.~P., {et~al.} 2010,
  \href{http://dx.doi.org/10.1016/j.newast.2009.12.002}{\JournalTitle{New
  Astron}, 15, 433}

\bibitem[{Mustill \& Villaver(2012)}]{Mustill2012}
Mustill, A.~J., \& Villaver, E. 2012,
  \href{http://dx.doi.org/10.1088/0004-637X/761/2/121}{\JournalTitle{Astrophys.
  J}, 761, 13}

\bibitem[{Mustill {et~al.}(2018)Mustill, Villaver, Veras, G{\"{a}}nsicke, \&
  Bonsor}]{Mustill2018}
Mustill, A.~J., Villaver, E., Veras, D., G{\"{a}}nsicke, B.~T., \& Bonsor, A.
  2018, \href{http://dx.doi.org/10.1093/MNRAS/STY446}{\JournalTitle{Mon. Not.
  R. Astron. Soc.}, 476, 3939}

\bibitem[{Nataf {et~al.}(2013)Nataf, Gould, Fouqu{\'{e}}, Gonzalez, Johnson,
  Skowron, Udalski, Szyma{\'{n}}ski, Kubiak, Pietrzy{\'{n}}ski,
  Soszy{\'{n}}ski, Ulaczyk, Wyrzykowski, \& Poleski}]{Nataf2013}
Nataf, D.~M., Gould, A., Fouqu{\'{e}}, P., {et~al.} 2013,
  \href{http://dx.doi.org/10.1088/0004-637X/769/2/88}{\JournalTitle{Astrophys.
  J}, 769, 88}

\bibitem[{Nordhaus \& Spiegel(2013)}]{Norhaus2013}
Nordhaus, J., \& Spiegel, D.~S. 2013,
  \href{http://dx.doi.org/10.1093/mnras/stt569}{\JournalTitle{Mon. Not. R.
  Astron. Soc.}, 432, 500}

\bibitem[{Penny {et~al.}(2019)Penny, {Scott Gaudi}, Kerins, Rattenbury, Mao,
  Robin, \& {Calchi Novati}}]{Penny2019}
Penny, M.~T., {Scott Gaudi}, B., Kerins, E., {et~al.} 2019,
  \href{http://dx.doi.org/10.3847/1538-4365/aafb69}{\JournalTitle{ApJS}, 241,
  3}

\bibitem[{Press {et~al.}(1992)Press, Teukolsky, Vetterling, \&
  Flannery}]{Press1992}
Press, W., Teukolsky, S., Vetterling, W., \& Flannery, B. 1992, {Numerical
  Recipes} (Cambridge, UK, Cambridge University Press)

\bibitem[{Schreiber {et~al.}(2019)Schreiber, G{\"{a}}nsicke, Toloza, Hernandez,
  \& Lagos}]{Schreiber2019}
Schreiber, M.~R., G{\"{a}}nsicke, B.~T., Toloza, O., Hernandez, M.~S., \&
  Lagos, F. 2019,
  \href{http://dx.doi.org/10.3847/2041-8213/ab42e2}{\JournalTitle{ApJL}, L4}

\bibitem[{Sigurdsson {et~al.}(2003)Sigurdsson, Richer, Hansen, Stairs, \&
  Thorsett}]{Sigurdsson2003}
Sigurdsson, S., Richer, H.~B., Hansen, B.~M., Stairs, I.~H., \& Thorsett, S.~E.
  2003,
  \href{http://dx.doi.org/10.1126/science.1086326}{\JournalTitle{Science}, 301,
  193}

\bibitem[{Suzuki {et~al.}(2016)Suzuki, Bennett, Sumi, Bond, Rogers, Abe,
  Asakura, Bhattacharya, Donachie, Freeman, Fukui, Hirao, Itow, Koshimoto, Li,
  Ling, Masuda, Matsubara, Muraki, Nagakane, Onishi, Oyokawa, Rattenbury,
  Saito, Sharan, Shibai, Sullivan, Tristram, \& Yonehara}]{Suzuki2016}
Suzuki, D., Bennett, D.~P., Sumi, T., {et~al.} 2016,
  \href{http://dx.doi.org/10.3847/1538-4357/833/2/145}{\JournalTitle{Astrophys.
  J}, 833, 145}

\bibitem[{Toonen {et~al.}(2017)Toonen, Hollands, G{\"{a}}nsicke, \&
  Boekholt}]{Toonen2017}
Toonen, S., Hollands, M., G{\"{a}}nsicke, B.~T., \& Boekholt, T. 2017,
  \href{http://dx.doi.org/10.1051/0004-6361/201629978}{\JournalTitle{Astron.
  Astrophys.}, 602, 16}

\bibitem[{Tremblay {et~al.}(2016)Tremblay, Cummings, Kalirai, G{\"{a}}nsicke,
  Gentile-Fusillo, \& Raddi}]{Tremblay2016}
Tremblay, P.~E., Cummings, J., Kalirai, J.~S., {et~al.} 2016,
  \href{http://dx.doi.org/10.1093/mnras/stw1447}{\JournalTitle{Mon. Not. R.
  Astron. Soc.}, 461, 2100}

\bibitem[{Vanderburg {et~al.}(2015)Vanderburg, Johnson, Rappaport, Bieryla,
  Irwin, Lewis, Kipping, Brown, Dufour, Ciardi, Angus, Schaefer, Latham,
  Charbonneau, Beichman, Eastman, McCrady, Wittenmyer, \&
  Wright}]{Vanderberg2015}
Vanderburg, A., Johnson, J.~A., Rappaport, S., {et~al.} 2015,
  \href{http://dx.doi.org/10.1038/nature15527}{\JournalTitle{Nature}, 526, 546}

\bibitem[{Vanderburg {et~al.}(2020)Vanderburg, Rappaport, Xu, Crossfield,
  Becker, Gary, Murgas, Blouin, Kaye, Palle, Melis, Morris, Kreidberg, Gorjian,
  Morley, Mann, Parviainen, Pearce, Newton, Carrillo, Zuckerman, Nelson,
  Zeimann, Brown, Tronsgaard, Klein, Ricker, Vanderspek, Latham, Seager, Winn,
  Jenkins, Adams, Benneke, Berardo, Buchhave, Caldwell, Christiansen, Collins,
  Col{\'{o}}n, Daylan, Doty, Doyle, Dragomir, Dressing, Dufour, Fukui, Glidden,
  Guerrero, Guo, Heng, Henriksen, Huang, Kaltenegger, Kane, Lewis, Lissauer,
  Morales, Narita, Pepper, Rose, Smith, Stassun, \& Yu}]{Vanderburg2020}
Vanderburg, A., Rappaport, S.~A., Xu, S., {et~al.} 2020,
  \href{http://dx.doi.org/10.1038/s41586-020-2713-y}{\JournalTitle{Nature},
  585, 363}

\bibitem[{Vandorou {et~al.}(2020)Vandorou, Bennett, Beaulieu, Alard, Blackman,
  Cole, Bhattacharya, Bond, Koshimoto, \& Marquette}]{Vandorou2020}
Vandorou, A., Bennett, D.~P., Beaulieu, J.-P., {et~al.} 2020,
  \href{http://dx.doi.org/10.3847/1538-3881/aba2d3}{\JournalTitle{Astrophys.
  J}, 160, 121}

\bibitem[{Veras(2016)}]{Veras2016}
Veras, D. 2016,
  \href{http://dx.doi.org/10.1098/rsos.150571}{\JournalTitle{Royal Society Open
  Science}, 3, 150571}

\bibitem[{Villaver \& Livio(2007)}]{Villaver2007}
Villaver, E., \& Livio, M. 2007,
  \href{http://dx.doi.org/10.1086/516746}{\JournalTitle{Astrophys. J}, 661,
  1192}

\bibitem[{Zhu {et~al.}(2014)Zhu, Gould, Penny, Mao, \& Gendron}]{Zhu2014}
Zhu, W., Gould, A., Penny, M., Mao, S., \& Gendron, R. 2014,
  \href{http://dx.doi.org/10.1088/0004-637X/794/1/53}{\JournalTitle{Astrophys.
  J}, 794, 53}

\end{thebibliography}

\section*{\textbf{Data availability}}
The KECK Observatory data used in this study are freely available on the Keck Observatory Archive (https://koa.ipac.caltech.edu/cgi-bin/KOA/nph-KOAlogin). Data from the VISTA Variables in the Via Lactea (VVV) survey are available on the European Southern Observatory archive (http://archive.eso.org/wdb/wdb/adp/phase3\_main/form?phase3\_collection=VVV\&release\_tag=6). Data used to model the light curve are available from the authors on request. 

\section*{\textbf{Code availability}}
The KECK pipeline is available on Github (https://github.com/blackmanjw/KeckPipeline). The Bayesian analysis code of D.P. Bennett uses routines from {\citep{Press1992}} which are subject to restricted availability.

\section*{\textbf{Acknowledgements}} \label{sec:ak} 
\indent Data presented in this work was obtained at the W. M. Keck Observatory from telescope time allocated to the National Aeronautics and Space Administration through the agencies scientific partnership with the California Institute of Technology and the University of California.  \textbf{Funding:} This work was supported by the University of Tasmania through the UTAS Foundation, ARC grant DP200101909 and the endowed Warren Chair in Astronomy. We acknowledge the support of ANR COLD WORLDS (ANR-18-CE31-0002) at the Institut d'astrophysique de Paris and the Laboratoire d'astrophysique de Bordeaux. D.P.B., A.B., N.K., C.R. and S.K.T. were supported by NASA through grant NASA-80NSSC18K0274 and by NASA award number 80GSFC17M0002. E.B. gratefully acknowledges support from NASA grant 80NSSC19K0291. Work by N.K. is supported by JSPS KAKENHI Grant Number JP18J00897. C.D. acknowledges financial support from the State Agency for Research of the Spanish MCIU through the ``Center of Excellence Severo Ochoa" award to the Instituto de Astrof\'isica de Andaluc\'ia (SEV-2017-0709), and the Group project Ref. PID2019-110689RB-I00/AEI/10.13039/501100011033. D.V. gratefully acknowledges the support of the STFC via an Ernest Rutherford Fellowship (grant ST/P003850/1).

\section*{\textbf{Author contributions}}
J.W.B. led the photometric and formal analysis and wrote the manuscript. J.W.B, V.B. and J.-P.B. took and reduced the photometric data using a pipeline written by J.W.B. and A.V. with contributions from J.-B.M. for magnitude calibrations. D.P.B., J.-P.B and A.A.C discussed conceptual and analysis approaches. D.P.B. was the PI of the KECK telescope proposal, led the planning of the observations, and conducted the light curve modeling and Bayesian analyses. C.D. provided insight into single and double white dwarf planetary systems. A.B. and E.B. assisted with proper motion calculations. A.B. and N.K assisted with observing on Keck. C.R. calculated the parallax, proper motion and lens prediction contours. C.A. worked on the PSF analysis and the determining of detection limits. D.P.B., J.-P.B., A.A.C., C.D., S.T. and D.V. contributed to the review and editing of the manuscript.

\section*{\textbf{Competing interests}}
The authors have no competing interests to declare.\\

Correspondence and requests for materials should be addressed to Joshua Blackman (joshua.blackman@utas.edu.au).\\

\end{document}